\documentclass[sigconf]{acmart}
\usepackage{float}
\newcommand{\myhl}[1]{\textcolor{black}{#1}}
\newcommand{\newhl}[1]{\textcolor{black}{#1}}
\newcommand{\nnhl}[1]{\textcolor{black}{#1}}
\usepackage{graphicx}
\usepackage{multirow}
\usepackage{footnote}
\usepackage{subfigure}
\usepackage{enumitem}

\usepackage[ruled,linesnumbered]{algorithm2e}
\AtBeginDocument{%
  }

\copyrightyear{2024}
\acmYear{2024}
\setcopyright{acmlicensed}\acmConference[CIKM '24]{Proceedings of the 33rd ACM International Conference on Information and Knowledge Management}{October 21--25, 2024}{Boise, ID, USA}
\acmBooktitle{Proceedings of the 33rd ACM International Conference on Information and Knowledge Management (CIKM '24), October 21--25, 2024, Boise, ID, USA}
\acmDOI{10.1145/3627673.3679681}
\acmISBN{979-8-4007-0436-9/24/10}

\begin{document}
\begin{sloppypar}
\title[Relative Contrastive Learning for Sequential Recommendation with Similarity-based Positive Pair Selection]{Relative Contrastive Learning for Sequential Recommendation with Similarity-based Positive Pair Selection}

\author{Zhikai Wang}
\affiliation{%
 \institution{DAMO Academy}
 \city{Hangzhou}
 \country{China}}
\email{wangzhikai.wzk@alibaba-inc.com}

\author{Yanyan Shen}
\affiliation{%
 \institution{Shanghai Jiao Tong University}
 \city{Shanghai}
 \country{China}}
\email{shenyy@sjtu.edu.cn}

\author{Zexi Zhang}
\affiliation{%
 \institution{Shanghai Jiao Tong University}
 \city{Shanghai}
 \country{China}}
\email{zhang-zexi@sjtu.edu.cn}

\author{Li He}
\affiliation{%
 \institution{Meituan}
 \city{Shanghai}
 \country{China}}
\email{heli18@meituan.com}

\author{Yichun Li}
\affiliation{%
 \institution{Meituan}
 \city{Shanghai}
 \country{China}}
\email{yichun.li@meituan.com}

\author{Hao Gu}
\affiliation{%
 \institution{Meituan}
 \city{Shanghai}
 \country{China}}
\email{guhao02@meituan.com}

\author{Yinghua Zhang}
\affiliation{%
 \institution{Meituan}
 \city{Shanghai}
 \country{China}}
\email{yzhangdx@outlook.com}
\renewcommand{\shortauthors}{Zhikai Wang et al.}


\begin{abstract}
Contrastive Learning~(CL) enhances the training of sequential recommendation~(SR) models through informative self-supervision signals. Existing methods often rely on data augmentation strategies to create positive samples and promote representation invariance. Some strategies such as item reordering and item substitution may inadvertently alter user intent. Supervised Contrastive Learning~(SCL) based methods find an alternative to augmentation-based CL methods by selecting same-target sequences~(interaction sequences with the same target item) to form positive samples. However, SCL-based methods suffer from the scarcity of same-target sequences and consequently lack enough signals for contrastive learning.
In this work, we propose to use similar sequences~(with different target items) as additional positive samples and introduce a \textbf{R}elative \textbf{C}ontrastive \textbf{L}earning (RCL) framework for sequential recommendation. RCL comprises a dual-tiered positive sample selection module and a relative contrastive learning module. The former module selects same-target sequences as strong positive samples and selects similar sequences as weak positive samples. The latter module employs a weighted relative contrastive loss, ensuring that each sequence is represented closer to its strong positive samples than its weak positive samples.
We apply RCL on two mainstream deep learning-based SR models, and our empirical results reveal that RCL can achieve 4.88\% improvement averagely than the state-of-the-art SR methods on five public datasets and one private dataset. The code can be found at https://github.com/Cloudcatcher888/RCL.
\end{abstract}


%

\maketitle

\section{Introduction}
\label{sec:introduction}
Sequential Recommendation models~\cite{MIND,C2Rec,FGNN,GAG,PosRec,time_lstm,din,fpmc,hrm,TAMIC,FGNN, GAG, PosRec,BERT4Rec,SASRec} predict a user's subsequent interaction based on their historical sequence, which aims at capturing both short-term preferences and long-term evolving interests. However, sequential recommendation models like GRU4Rec~\cite{GRU4Rec} and SASRec~\cite{SASRec} encounter challenges of the inherent sparsity and noise within the data. In response, Contrastive Learning~(CL) based methods~\cite{MCLRec,CL4SRec, CoSeRec, TiCoSeRec}, exemplified by recent advancements~\cite{SimCLR,SimSiam,DIM}, leverage diverse views to enhance the learned representations of sequences, offering a potential solution to address these data-related limitations in sequential recommendation models.

While CL aims to enhance sequence representations by maximizing agreement among augmented views of the same sequence and distancing views of different sequences, current CL-based methods~\cite{CL4SRec,CT4Rec,crossdomain,S3Rec, DHCN, MHCN, MMInfoRec} often rely on instinctual identification of random augmentation operations, like random sequence or model perturbations (`crop' and `mask' operations). The process of identifying effective augmentation operations requires specific domain knowledge and meticulous design, potentially causing interference with the original user intent and introducing additional noise during model training~\cite{DuoRec,relevant}.


\begin{figure}[t]
    \centering
    \includegraphics[width=0.49\linewidth]{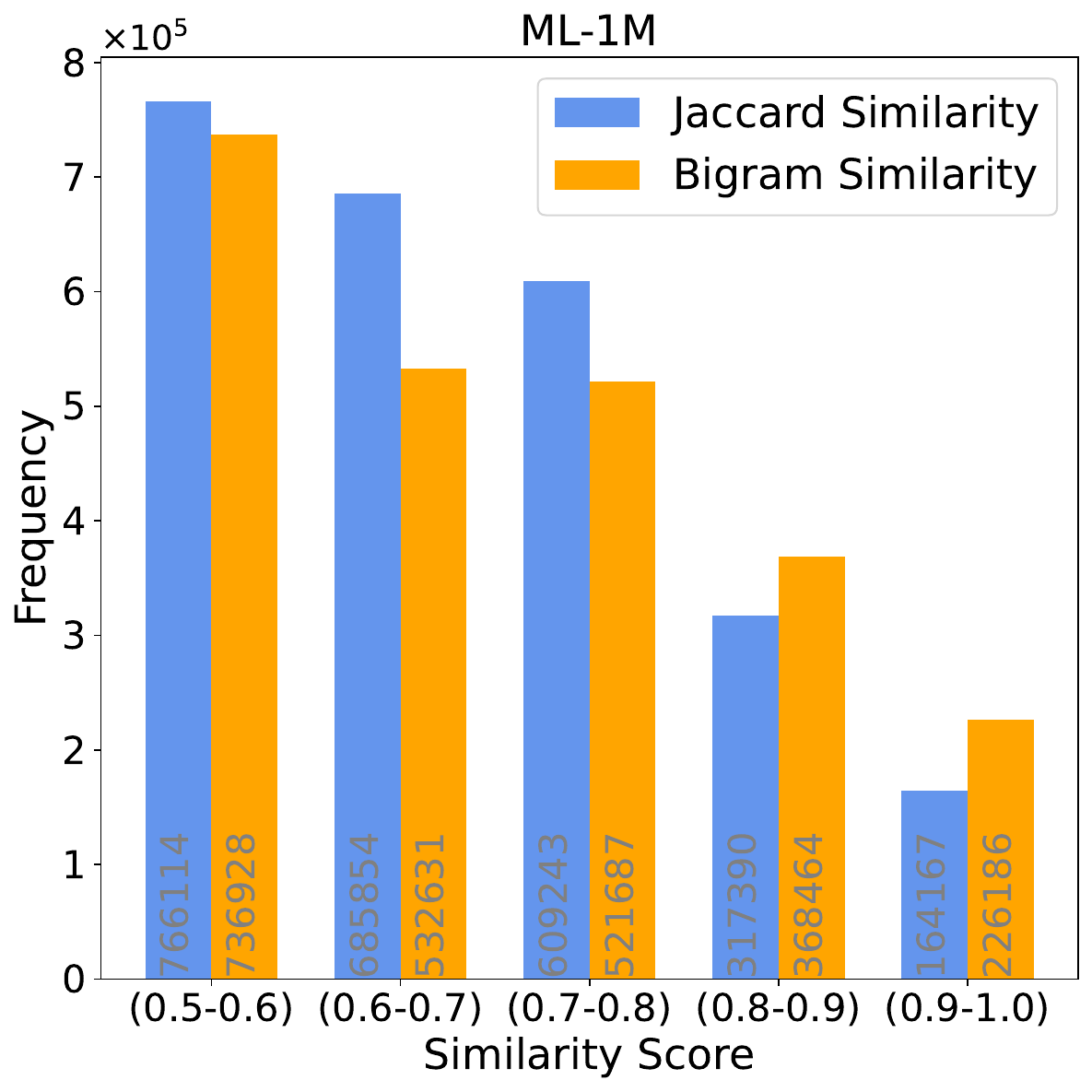}
    \includegraphics[width=0.49\linewidth]{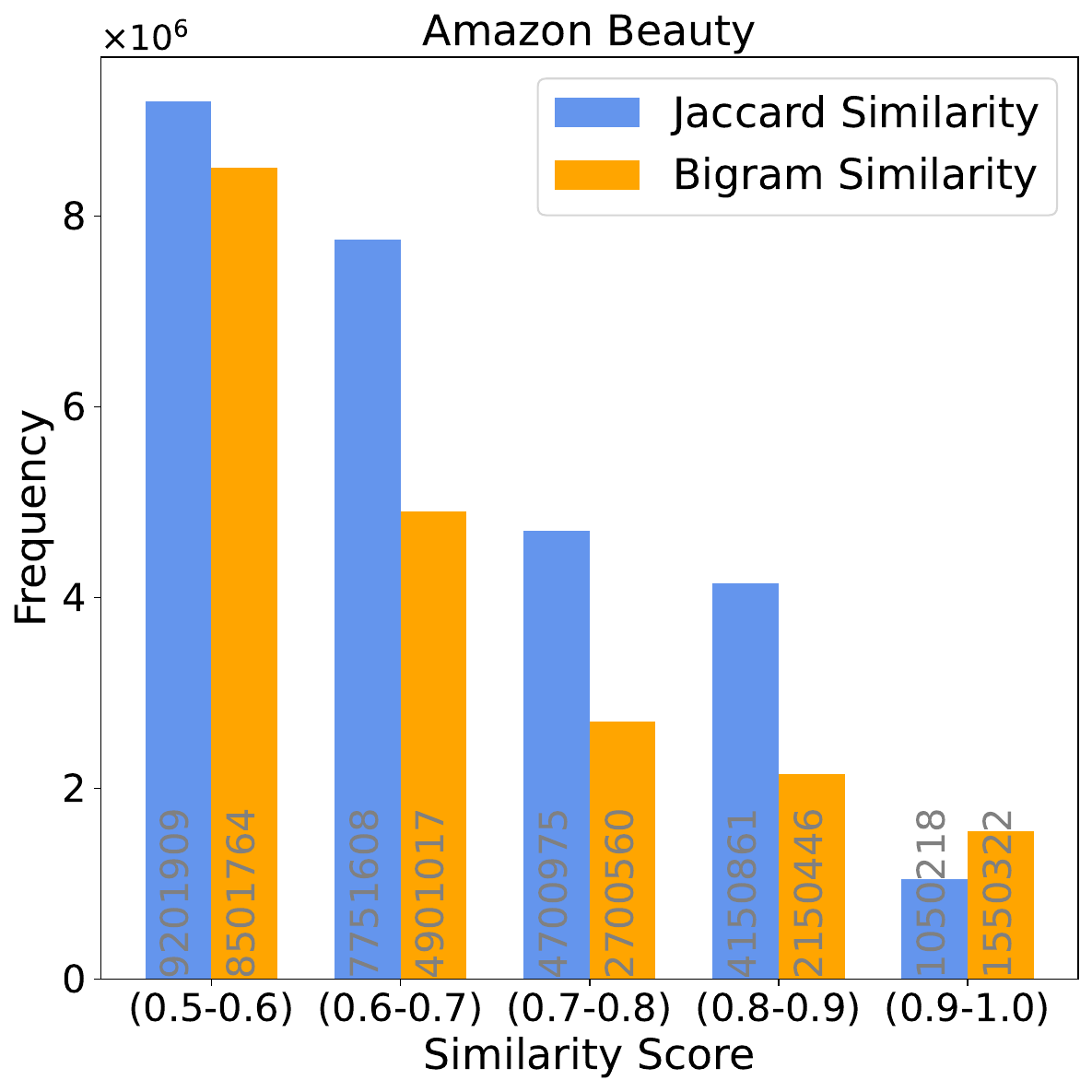}
    \caption{Similarity frequency histograms of sequence pairs with different target items on ML-1M and Amazon Beauty datasets.}
    \label{fig:sfh}
\end{figure}
 Recently, supervised contrastive learning~(SCL) based methods~\cite{DuoRec,HPM,ContraRec} 
 suggest choosing same-target sequences~(interaction sequences with the same target item) as positive samples. They leverage real sequences instead of augmented sequences and aim at improving the alignment among sequences sharing the same intents. For instance, ContraRec~\cite{ContraRec} introduces a contrastive learning task to encourage same-target sequences to have similar representations, called context-context contrast.  In a similar vein, DuoRec~\cite{DuoRec} argues same-target sequences natually contain the same intents and designs a supervised contrastive loss to draw the representations of same-target sequences closer. However, these approaches encounter limitations due to the scarcity of same-target sequences. In the ML-1M and Amazon Beauty datasets, 47.5\% and 52.7\% of the sequences do not have any same-target sequences, respectively. Consequently, SCL based methods cannot obtain supervised contrast signals on these sequences with no same-target sequences, leading to limited performance improvement.



   
In this paper, we propose to treat similar sequences~(with different target items) as additional positive samples, which has two reasons.
First, a certarin proportion of sequences with different target items actually exhibit a considerable degree of similarity, the representations of which should be naturally close in the latent feature space. 
We provide the similarity frequency histograms of sequence pairs with different target items of ML-1M and Amazon Beauty in Figure~\ref{fig:sfh}. We use the Jaccard and Bigram Similarity~\cite{Bigram}~(two adjacent items are used as one gram) as the sequence similarity metrics. 
We find top 5.31\%/4.28\% of sequence pairs have Jaccard Similarity larger than 0.7 on ML-1M/Amazon Beauty, each pair of which can be regarded as similar sequences and naturally should have close representations. 
  Second, similar sequences can serve as a supplement to the positive samples with the same target item. In ML-1M and Amazon Beauty, 100\% of the sequences have at least 319/744 similar sequences~(Jaccard similarity is larger than 0.7), ensuring that all sequences have corresponding positive samples. Though treating similar sequences as additional positive samples is beneficial, it is challenging to choose suitable similarity metrics and an effective sampling strategy to select similar sequences that truly share the same intents like same-target sequences.

It is further required to consider how to use both same-target sequences and similar sequences as positive samples. 
  Intuitively, the target item directly reflects a user's future intent, while an interaction sequence represents the user's past intent, which serves as the indirect reflection of his/her future intent. Consequently, we give precedence to same-target sequences as strong positive samples, while treating similar sequences as weak positive samples. However, it is also challenging to design an appropriate contrastive learning loss function to ensures that each sequence is represented closer to its strong positive samples than its weak positive samples. If we directly use an infoNCE loss and treat strong/weak positive samples as the numerator/denominator of the infoNCE loss, the loss will inevitably push the representations of weak positive samples too far away from the center sequence.


To tackle the two challenges mentioned above, we introduce a \textbf{R}elative \textbf{C}ontrastive \textbf{L}earning~(RCL) method for sequential recommendation that incorporates a dual-tiered positive sample selection module and a relative contrastive learning module. 
Specifically, the positive sample selection module resolves the first challenge, which introduces a weighted sequence sampling strategy based on different order sensitive/insensitive similarity metrics to select similar sequences truly sharing the same intents. 
In conjunction with this, the relative contrastive learning module utilizes a weighted relative contrastive loss to overcome the second challenge, which adds a loss boundary on general infoNCE loss to effectively control the relative loss magnitudes of strong positive samples and weak positive samples. Extensive experiments on six real datasets demonstrate the effectiveness of the proposed RCL framework in terms of the recommendation performance. 

The main contributions of this paper are summarized as follows.
\begin{itemize}[leftmargin=10pt, rightmargin=0pt]
    \item We present a Relative Contrastive Learning~(RCL) based method for sequential recommendation, which uses both same-target sequences and similar sequences as positive samples for contrative learning to further improve the alignment among sequences sharing the same intents than SCL based methods.
    \item A dual-tiered positive sample selection module is proposed, which introduces a weighted sampling strategy based on several similarity metrics to select similar sequences truly sharing the same intents.
    \item A relative contrastive learning module is proposed to control the relative loss magnitudes of strong positive samples and weak positive samples, ensuring that each sequence is represented closer to its strong positive samples than its weak positive samples.
    \item We validate our RCL method via comprehensive experimentation on two sequential recommendation models, i.e., SASRec and FMLP~\cite{fmlp}. The empirical outcomes show that our approach achieves 4.88\% improvement averagely against state-of-the-art methods across five public and one private datasets.
\end{itemize}
   

\begin{figure*}[h]
    \centering
    \includegraphics[width=0.96\linewidth]{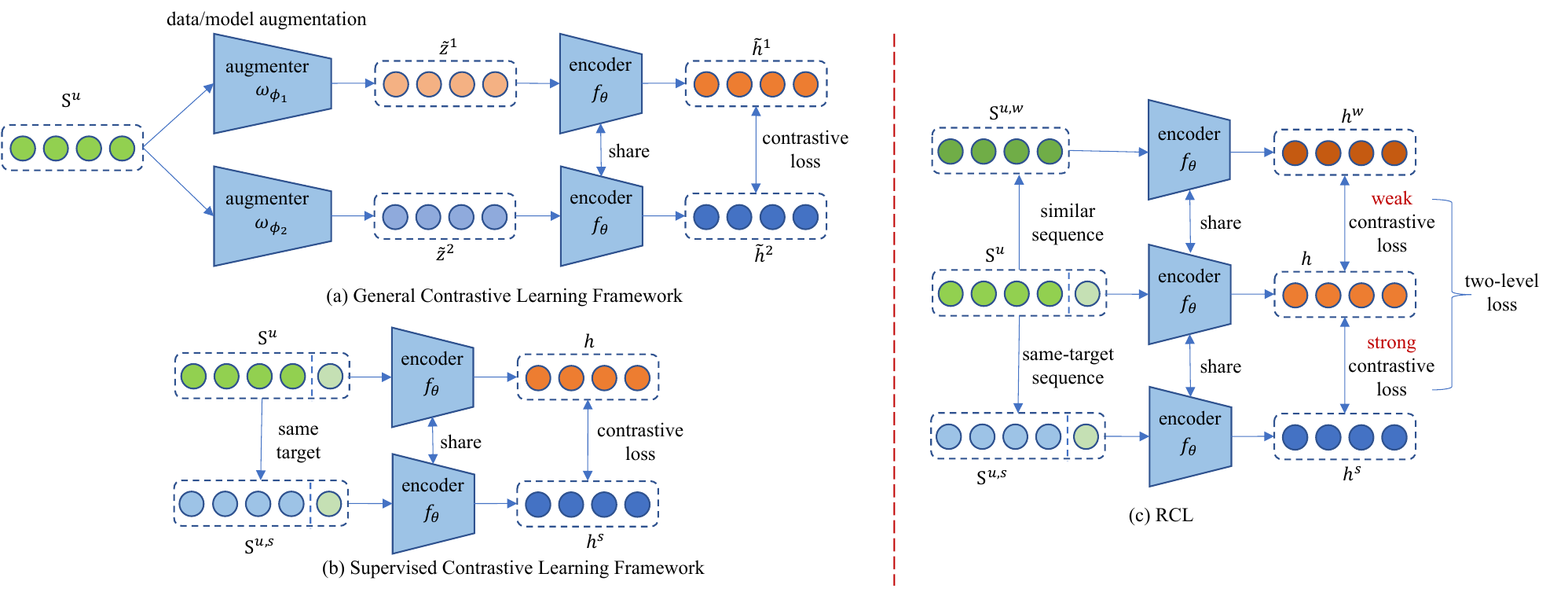}
    \caption{In the General Contrastive Learning Framework~(a), the typical components include a data or model based augmentation module, a user representation encoder, and a contrastive loss function.
In the Supervised Contrastive Learning Framework~(b), the augmentation module is substituted with a randomly sampled same-target positive.
The proposed RCL~(c) differs by employing a dual-tiered positive pair selection module, which treats same-target sequences as strong positive samples and treats similar sequences as weak positive samples. A relative contrastive learning module is employed to manage the dual-tiered positive samples.}
    \label{fig:overview}
\end{figure*}

\section{Preliminary}
\subsection{Problem Definition}
Sequential Recommendation (SR) aims to suggest the next item that a user is likely to interact with, leveraging their historical interaction data. Assuming that user sets and item sets are $\mathcal{U}$ and $\mathcal{V}$ respectively, user $u \in \mathcal{U}$ has a sequence of interacted items $S_u=\left\{v_{1,u}, \ldots, v_{{\left|S_u\right|},u}\right\}$. $v_{i,u} \in \mathcal{V}\left(1 \leq i \leq\left|S_u\right|\right)$ represents an interacted item at position $i$ of user $u$ within the sequence, where $\left|S_u\right|$ denotes the sequence length. Given the historical interactions $S_u$, the goal of SR is to recommend an item from the set of items $\mathcal{V}$ that the user $u$ may interact with at step $\left|S_u\right|+1$:
\begin{equation}
\arg \max _{v' \in \mathcal{V}} P\left(v_{{\left|S_u\right|+1},u}=v' \mid S_u\right).
\end{equation}

\subsection{Sequential Recommendation Model}
Our method incorporates a backbone SR model comprising three key components: (1) an embedding layer, (2) a representation learning layer, and (3) a next item prediction layer.
\subsubsection{Embedding Layer}
Initially, the entire item set $\mathcal{V}$ is embedded into a shared space, resulting in the creation of the item embedding matrix $\mathbf{M} \in \mathbb{R}^{|\mathcal{V}| \times d}$. Given an input sequence $S_u$, the sequence's embedding $\mathbf{E}_u \in \mathbb{R}^{\left|S_u\right| \times d}$ is initialized, and $\mathbf{E}_u$ is defined as $\mathbf{E}_u=\left\{\mathbf{m}_{1}\oplus\mathbf{p}_1, \mathbf{m}_{2}\oplus\mathbf{p}_2, \ldots, \mathbf{m}_{\left|S_u\right|}\oplus\mathbf{p}_{\left|S_u\right|}\right\}$. Here, $\mathbf{m}_{i} \in \mathbb{R}^d$ represents the embedding of the item at position $i$ in the sequence, $\mathbf{p}_i \in \mathbb{R}^d$ signifies the positional embedding within the sequence, $\oplus$ denotes the element-wise addition, and $n$ denotes the sequence's length.
\subsubsection{Representation Learning Layer}
Given the sequence embedding $\mathbf{E}_u$, a deep neural encoder denoted as $f_\theta(\cdot)$ is utilized to learn the representation of the sequence. The output representation is calculated as:
\begin{equation}
\mathbf{h}_u=f_\theta\left(\mathbf{E}_u\right)\in \mathbb{R}^d.
\end{equation}


Finally, the predicted interaction probability of each item can be calculated as: 
\begin{equation}
  \hat{\mathbf{y}}=\operatorname{softmax}\left(\mathbf{h}_{u} \mathbf{M}^{\top}\right)\in \mathbb{R}^{|\mathcal{V}|}.
\end{equation}
The training loss used for optimizing the sequential recommendation model is as follows:
\begin{equation}
	\mathcal{L}^{rec}=-\sum_{u \in \mathcal{U}}\log \hat{\mathbf{y}}[v_{|S_u|+1,u}],\label{eq:rec}
\end{equation}
where $v_{|S_u|+1,u}$ denotes the ground-truth target item of user $u$.


\section{Methodology}
\label{sec:methodology}

As depicted in Figure~\ref{fig:overview}(a), the general contrastive learning framework typically comprises a stochastic augmentation module, a user representation encoder, and a contrastive loss function. The supervised contrastive learning framework, as illustrated in Figure~\ref{fig:overview}(b), discards the stochastic augmentation module and chooses same-target sequences as positive samples.
The proposed RCL, as shown in Figure~\ref{fig:overview}(c), employs same-target sequences as strong positive samples and similar sequences as weak positive samples for each sequence, respectively. RCL implements a dual-tiered positive sample selection module, effectively capturing inherent similarities in user intent across sequences. Furthermore, RCL introduces a weighted relative contrastive learning module to ensure that each sequence is represented closer to its strong positive samples than its weak positive samples.

\subsection{Dual-tiered Positive Sample Selection}
In this section, we delve into RCL's method for selecting positive samples. The SCL-based
 methods~\cite{HPM,DuoRec} have cautioned against the use of data or model-based augmentation in positive sample selection, which could potentially disrupt the inherent user intents. Instead, they try to cluster sequences with identical intents by using the same target sequences as positive samples in contrastive learning paradigm. However, these methods encounter limitations due to the scarcity of same-target sequences. We observe that some negative samples sampled by SCL-based method~\cite{DuoRec} in each minibatch actually have high Jaccard or Bigram similarity. Intuitively, the representations of the sequences with high similarity should be close in the latent feature space. Treating the highly similar sequences as negative samples may incorrectly push the representations away and hence degrade the learned representations. Instead, these similar sequences can be used as positive samples to enlarge the original positive sample set. 

Furthermore, considering the interaction sequence as a representation of a user's historical preferences and the target item as an indicator of their future interaction intent, we prioritize treating same-target sequences as strong positive samples ($\mathcal{SP}_u$) and similar sequences as weak positive samples ($\mathcal{WP}_u$) for each sequence $S_u$. 
In what below, we will discuss how we select strong positive samples and weak positive samples.

\subsubsection{Same-target Strong Positive Sample Selection}
In this section, we will introduce how RCL selects the same-target sequences as strong positive samples briefly. For each sequence $S_u$, its strong positive samples can be formally defined as: \begin{equation}
  \mathcal{SP}_u=\{S_{u,a}|v_{{\left|S_{u,a}\right|+1},a}=v_{{\left|S_u\right|+1},u}\},
\end{equation}
where $S_{u,a}$ is the sequence sharing the same target item with $S_u$.
 Following the SCL-based methods~\cite{DuoRec,HPM}, RCL randomly picks one sequence $S_{u,a}$ from $\mathcal{SP}_u$ per iteration if $\mathcal{SP}_u$ is not empty.

\subsubsection{Similarity-based Weak Positive Pair Selection}
\label{sec:weak}
In this section, we will discuss how RCL selects similar sequences as weak positive samples $\mathcal{WP}_u$ for sequence $S_u$. 
We first introduce several similarity metrics to assess the relationship between any two user interaction sequences, $S_{u_1}$ and $S_{u_2}$. 

%
%
\begin{itemize}[leftmargin=10pt,rightmargin=0pt]
	\item \textit{Jaccard Similarity:} 
	The Jaccard Similarity between two interaction sequences can be calculated as follows:
	\begin{equation}
		s_{u_1,u_2}=\text{Jaccard Similarity} (A_{u_1}, A_{u_2}) = \frac{|A_{u_1} \cap A_{u_2}|}{|A_{u_1} \cup A_{u_2}|},
	\end{equation}
	where \(A_{u_1}\) represents the set of unique elements in the first interaction sequence $S_{u_1}$. \(A_{u_2}\) represents the set of unique elements in the second interaction sequence $S_{u_2}$. 
	The Jaccard Similarity measures the proportion of common elements between the sequences relative to the total unique elements in both sequences.
	
	\item \textit{TF-IDF:}
	Drawing inspiration from text analysis, we conceptualize interaction sequences as sentences and items as individual words. Then the weight for each item $v$ in sequence $S_u$ is $w_{v,u}=tf_{v,u}\log\left(\frac{N}{df_v}\right)$, where $tf_{v,u}$ is the frequency of $v$ in $u$ and $df_v$ is the frequency of $v$ over the whole dataset.
	Each sequence can be transformed as item weight vector $\mathbf{w}_u=[w_{1,u}, \cdots, w_{|\mathcal{V}|,u}]$. We calculate the cosine similarity between each two vectors $\mathbf{w}_{u_1}$ and $\mathbf{w}_{u_2}$ as the sequence similarity: 
	\begin{equation}
		s_{u_1,u_2}=\frac{\mathbf{w}_{u_1}^\top \mathbf{w}_{u_2}}{|\mathbf{w}_{u_1}||\mathbf{w}_{u_2}|}.
	\end{equation}
	\item \textit{N-grams Similarity:} Compared to Jaccard Similarity, it considers the shared subsequences (N-grams) of a specific length~(n). Formally, the N-grams similarity can be calculated as follows:
	\begin{equation}
		s_{u_1,u_2}=\frac{|A_{u_1,ngram} \cap A_{u_2,ngram}|}{|A_{u_1,ngram} \cup A_{u_2,ngram}|},
	\end{equation}
	where $A_{u_1,ngram}$ and $A_{u_2,ngram}$ represent the sets of unique N-grams in interaction sequence $S_{u_1}$ and $S_{u_2}$, respectively.
	\item \textit{Levenshtein Distance:} This similarity metric refers to the minimum number of insertions, deletions, and substitutions required to transform one sequence into the other, which measures the dissimilarity between the sequences $S_{u_1}$ and $S_{u_2}$, where a smaller distance indicates greater similarity:
\[d(S_{u_1}, S_{u_2}) = \min \begin{cases}
d(S_{u_1}', S_{u_2}') + c(S_{u_1}[m], S_{u_2}[n]) \\
d(S_{u_1}, S_{u_2}') + 1 \\
d(S_{u_1}', S_{u_2}) + 1,
\end{cases}\]
where \(S_{u_{1/2}}'\) are the string \(S_{u_{1/2}}\) with their last characters removed.
\(m\) and \(n\) are the lengths of \(S_{u_{1/2}}\).
\(c(x, y)\) is a function that returns \(0\) if characters \(x\) and \(y\) are the same, and \(1\) otherwise.

	\item \textit{Semantic Distance:} This similarity metric refers to the cosine similarity of sequence representations:
	\begin{equation}
		s_{u_1,u_2}=\frac{\mathbf{h}_{u_1}^\top \mathbf{h}_{u_2}}{|\mathbf{h}_{u_1}||\mathbf{h}_{u_2}|},
	\end{equation}
	where $\mathbf{h}_{u_1}$ and $\mathbf{h}_{u_2}$ denote the representations of the interaction sequences $S_{u_1}$ and $S_{u_2}$.
\end{itemize}
\textit{Jaccard Similarity} and \textit{TF-IDF} both disregard order, focusing solely on the count of interaction items. \textit{N-grams Similarity}, on the other hand, is sensitive to local order, with larger N values increasing this order sensitivity. Meanwhile, \textit{Levenshtein Distance} and \textit{Semantic Distance} are explicitly sensitive to the order of elements within the sequences.
The similarity calculation methods are not limited to the aforementioned five types. \myhl{Our primary innovation is not the design of similarity metrics but rather in effectively leveraging weak positives for contrastive learning. We believe that the performance on the basic similarity metrics can better validate the advantages of the overall method.}
The choice of similarity metrics is examined in the ablation study~(Section~\ref{sec:ablation}).

Then we aim at selecting weak positive samples for each sequence $S_u$ using one similarity metric provided above. A direct strategy is choosing the sequence $S_{u,b}$ with the highest similarity score as the weak positive sample. 
However, the selected positive samples remain fixed across different epochs and other sequences with high similarity are not utilized for training.
To refine this strategy, for each interaction sequence $S_u$, we now select all sequences with the top $\alpha$ highest similarity scores as weak positive samples $\mathcal{WP}_{u}$. Here, $\alpha$ is a hyper-parameter. We randomly choose one sequence $S_{u,b}$ from $\mathcal{WP}_{u}$ in each iteration during training. Intuitively, a sequence with higher similarity is preferred to be chosen. Consequently, the sampling probability for each sequence $S_{u,b}$ in $\mathcal{WP}_u$ is set to be proportional to the similarity score, which is formally defined as: 
\begin{equation} 
p_b = \frac{s_{u,b}}{\sum_{c\in\mathcal{WP}_u} s_{u,c}}, 
\end{equation} 
where $s_{u,b}$ is the similarity score between sequences $S_u$ and $S_{u,b}$, and the denominator aggregates the similarity scores of all positives in the weak positive set of sequence $S_u$.



\subsection{Relative Contrastive Learning}
\label{sec:loss}
In this section, we propose a relative contrastive learning module aimed at ensuring that each sequence is represented closer to its strong positive samples than its weak positive samples.
Generally, for a sequence $S_u$~(referred to as center sequence), we aim to satisfy the condition for $S_{u,a}\in \mathcal{SP}_{u}$ and $S_{u,b}\in \mathcal{WP}_{u}$:
\begin{equation}
    \mathbf{h}_u^\top \mathbf{h}_a > \mathbf{h}_u^\top \mathbf{h}_b,
\end{equation}
where $\mathbf{h}_u$,$\mathbf{h}_a$, and $\mathbf{h}_b$ are the representations of sequences $S_u$, $S_{u,a}$, and $S_{u,b}$, respectively.

We propose to use an infoNCE-based loss function~\cite{UATL} for contrastive learning, which is originally designed to control the relative magnitudes of losses across different hierarchical levels of labels. Although in our scenario the labels do not overlap among different sequences, the concept of regulating the relative sizes of losses is still applicable. We therefore categorize strong positive samples as high-level labels and weak positive samples as low-level labels. To ensure that the similarity score $\mathbf{h}_u^\top \mathbf{h}_a$ is greater than the score $\mathbf{h}_u^\top \mathbf{h}_b$, we implement a constraint on the loss associated with weak positive samples.
Specifically, we define the loss between $S_u$ and $S_{u,a}$ as:
\begin{equation}
    \label{eq:infonce}
    \mathcal{L}^{\text{pair}}(S_u, S_{u,a}) = -\log \frac{\exp \left(\mathbf{h}_u \cdot \mathbf{h}_a / \tau\right)}{\sum_{S_c \in A \backslash\{S_u\}} \exp \left(\mathbf{h}_u \cdot \mathbf{h}_c / \tau\right)},
\end{equation}
where $A$ denotes all sequences in a batch and $\tau$ denotes the temperature parameter.
The maximum loss (the largest distance) within strong positive samples $\mathcal{SP}_{u}$ is denoted as:
\begin{equation}
    \mathcal{L}^{max}_{u} = \max_{S_{u,a} \in \mathcal{SP}_{u}} L^{\text{pair}}(S_u, S_{u,a}).
\end{equation}
Consequently, the total loss for both strong positive samples and weak positive samples becomes:
\begin{equation}
    \label{eq:unweight}
    \begin{split}
    \mathcal{L}^{RCL} =  &\sum_{u \in \mathcal{U}}\mathcal{L}^{\text{pair}}(S_u, S_{u,a}) \\
    +&\sum_{u \in \mathcal{U}} \max \left(\mathcal{L}^{\text{pair}}(S_u, S_{u,b}), \mathcal{L}^{max}_{u}\right).
    \end{split}
\end{equation}
\myhl{The second term ensures that if the loss of any weak positive pair $L^{\text{pair}}(S_u,S_{u,b})$ is smaller than the largest loss among strong positive samples $L^{max}_{u}$ (the boundary), the gradient will pass through $L^{max}_{u}$ and push the boundary closer to sequence $S_u$. 
By controlling the upper bound of $\mathcal{L}^{\text{pair}}(S_u, S_{u,b})$, we ensure that all pairwise distances are collectively pushed towards the center, allowing better control over the contrastive objective. }
If the set of strong positive samples $\mathcal{SP}_u$ is empty, $L^{RCL}$ will solely consider weak positive samples and is simplified to:
\begin{equation}
\label{eq:weak}
\mathcal{L}^{RCL} = \sum_{u \in \mathcal{U}} \mathcal{L}^{\text{pair}}(S_u, S_{u,b}).
\end{equation}
In SCL methods~\cite{DuoRec,ContraRec}, augmentation is still necessary in case there are no positive samples with the same target item. However, in our framework, this scenario can be prevented.

\subsubsection{Similarity-Based Re-weighting}
Intuitively, higher similarity between sequences correspond to higher confidence in contrastive learning, thus necessitating a higher weight in the loss function for both positive samples and negative samples. To achieve this, we modify the loss between $S_u$ and $S_{u,a}$ into a weighted version:
\begin{equation}
\label{eq:weight}
    \mathcal{L}^{\text{weighted}}(S_u, S_{u,a}) = -\log \frac{s_{u,a}\exp \left(\mathbf{h}_u \cdot \mathbf{h}_a / \tau\right)}{\sum_{S_c \in A \backslash\{S_u\}} s_{u,c}\exp \left(\mathbf{h}_u \cdot \mathbf{h}_c / \tau\right)},
\end{equation}
where $s_{u,a}$ denotes the similarity score between $S_u$ and $S_{u,a}$.
Thus $L^{RCL}$ will be updated as:
\begin{equation}
    \label{eq:RCL}
    \begin{split}
    \mathcal{L}^{wRCL} =  &\sum_{u \in \mathcal{U}}\mathcal{L}^{\text{pair}}(S_u, S_{u,a}) \\
    +&\sum_{u \in \mathcal{U}} \max \left(\mathcal{L}^{\text{weighted}}(S_u, S_{u,b}), \mathcal{L}^{max}_{u}\right).
    \end{split}
\end{equation}
The overall objective of the joint learning is defined as:
\begin{equation}
\label{eq:total}
    \mathcal{L}^{total} = \mathcal{L}^{rec} + \lambda \mathcal{L}^{wRCL},
\end{equation}
where $\lambda$ is a hyper-parameter used to control the magnitude of the relative contrastive loss. We use $\mathcal{L}^{wRCL}$ as the default in performance comparison. The whole training algorithm is in Algorithm~\ref{alg:1}.

\begin{savenotes}
	\begin{algorithm}
		\caption{The RCL Algorithm}\label{alg:1}
		\KwIn{Training dataset $\left\{S_u\right\}_{u=1}^{|\mathcal{U}|}$, hyper-parameters $\lambda$,$\alpha$}
		\KwOut{Recommendation lists}
		Calculate the similarities between any two sequences in $\left\{S_u\right\}_{u=1}^{|\mathcal{U}|}$\;
		Collect strong positive samples $\mathcal{SP}_u$ and weak positive samples $\mathcal{WP}_u$ for each sequence $S_u$\footnote{If Semantic Distance is selected as the similarity metric, the sequence similarity becomes dynamically adaptive, requiring recalculation in each iteration to capture its evolving nature.}\;
		\For{$each~minibatch$}{
			Calculate the recommendation loss by Eq.~(\ref{eq:rec})\;
			Calculate the RCL loss by Eq.~(\ref{eq:RCL})\;
			Calculate the joint loss by Eq.~(\ref{eq:total})\;
			Jointly optimize the overall objective\;
		}
	\end{algorithm}
\end{savenotes}

\subsection{Complexity Analysis}
\label{timecost}
RCL introduces no additional parameters beyond generalized contrastive learning in Figure~\ref{fig:overview}\nnhl{(a)}, which involves $|\mathcal{V}|\times d+|\theta|$ parameters. $d$ and $\theta$ denote the embedding size and parameters of the sequence representation encoder.
\myhl{Given the total sequence count \( N \), the sequence similarity calculation can be performed in parallel using a GPU, where each thread of the GPU is responsible for calculating one sequence's similarities with all other sequences. The GPU memory only needs to store \( \alpha N^2 \) similarities (along with their corresponding sequence IDs), which will be replaced by newly calculated similarities with larger values.
The sequence similarity calculation has a time complexity of \( O\left(N^2 \times \frac{T(s)}{R}\right) \), where \( T(s) \) refers to the time cost for one sequence pair. \( R \) is the number of threads in the GPU, which is a large value. For \textit{Semantic Distance}, \( T(s) \) is \( d^2 \), where \( d \) is the embedding and hidden vector dimension. For the other four metrics, \( T(s) \) is \( \overline{|S|} \), where \( \overline{|S|} \) is the average sequence length. The GPU memory cost of sequence similarity calculation is \( \alpha N^2 \), where \( \alpha \) is \nnhl{the similarity threshold} and is set to a small value.} In practical recommender systems, narrowing down the scope to calculate the sequence similarity within a specific city or day can further decrease the complexity.


\begin{table}[t]
    \centering
    \caption{Statistics of the datasets after preprocessing.}
    \setlength{\tabcolsep}{2pt}
    \begin{tabular}{l|rrrrr}
    \toprule
    & \multirow{2}{*}{\# Users} & \multirow{2}{*}{\# Items} & \# Avg.  & \multirow{2}{*}{\# Actions} & \multirow{2}{*}{Sparsity} \\
    &          &          & Length               &            &          \\
    \midrule
    Beauty & 22,363 & 12,101 & 8.9 & 198,502 & $99.93 \%$ \\
    Sports & 35,598 & 18,357 & 8.3 & 296,337 & $99.95 \%$ \\
    ML-1M & 6,041 & 3,417 & 165.5 & 999,611 & $95.16 \%$ \\
    ML-20M & 138,493 & 27,278 & 144.4 & 20,000,263  & $99.47 \%$ \\
    Yelp & 30,499 & 20,068 & 10.4 & 317,182 & $99.95 \%$ \\
    \midrule
    Life & \multirow{2}{*}{2,508,449} & \multirow{2}{*}{276,331} &\multirow{2}{*}{40.8} & \multirow{2}{*}{102,399,201} & \multirow{2}{*}{$99.99 \%$}\\
    Service & & & & & \\
    \bottomrule
    \end{tabular}
    \label{tab:stats}
\end{table}

\section{Experiments}
\label{sec:experiments}
In experiments, we will answer the following research questions:
\begin{itemize}
\item \textbf{RQ1} How does the RCL perform compared with the state-of-the-art methods?
\item \textbf{RQ2} How does each component of the RCL contribute to its effectiveness? 
\item \textbf{RQ3} How do hyperparameters influence the performance of RCL?
\item \textbf{RQ4} Where do the improvements of the RCL come from?
\item \textbf{RQ5} How does RCL perform on the online sequential recommendation platform?
\end{itemize}

\subsection{Setup}
\subsubsection{Datasets}
The experiments cover six benchmark datasets, detailed in Table~\ref{tab:stats} after preprocessing:
\begin{itemize}[leftmargin=10pt, rightmargin=0pt]
    \item \textbf{Amazon Beauty} and \textbf{Sports}\footnote{https://jmcauley.ucsd.edu/data/amazon/} utilize the widely-used Amazon dataset with two sub-categories as previous baselines.
\item \textbf{MovieLens-1M/20M} (ML-1M/20M)\footnote{https://grouplens.org/datasets/movielens/1m/} are two versions of a popular movie recommendation dataset with different sizes.
\item \textbf{Yelp}\footnote{https://www.yelp.com/dataset} is widely used for business recommendation. Similar to previous works~\cite{DuoRec,MCLRec}, the interaction records after Jan. 1st, 2019 are used in our experiments.
\item \textbf{Life Service} \newhl{is a private sequential recommendation dataset, which is collected from Meituan Dianping platform for local life services such as restaurants and entertainment places. We focus on user interactions within a single day in a specific city. }
\end{itemize}

\myhl{We follow the settings of previous works~\cite{DuoRec,MCLRec,HPM}.
All interactions are treated as implicit feedback, filtering out users or items appearing fewer than five times. The maximum sequence length for the ML-1M/20M datasets is set to 200, whereas for the other four datasets, the maximum sequence length is set to 50.}

For evaluation, we use Top-$K$ Hit Ratio (HR@K) and Top-$K$ Normalized Discounted Cumulative Gain (NDCG@K) across $K$ values of $\{5,10\}$, evaluating rankings across the entire item set for fair comparison, following established methodologies~\cite{DuoRec}.

\begin{table*}[htpb]
    \centering
    \caption{Overall performance. Bold scores represent the highest results of all methods. Underlined scores stand for the highest results from previous methods. The RCL achieves the state-of-the-art result among all baseline methods.}
\begin{tabular}{c|c|ccccccc|c|c}
\toprule Dataset 
&	Metric	&	SASRec	&	CL4SRec	&	CT4Rec	&	MCLRec	&	DuoRec	&    ContraRec	&	HPM	&	RCL	&	Improv.	\\\midrule\multirow{4}{*}{Beauty}
																				
&	HR@5	&	0.0365 	&	0.0401 	&	0.0575 	&	0.0581 	&	0.0546 	&	0.0551 	&$\underline{	0.0572 	}$&$\mathbf{	0.0601 	}\pm	0.0011 	$&$	5.07 	\%$\\
&	HR@10	&	0.0627 	&	0.0683 	&	0.0856 	&	$\underline{0.0861}$ 	&	0.0845 	&	0.0855 	&	0.0860 	&$\mathbf{	0.0898 	}\pm	0.0005 	$&$	3.10 	\%$\\
&	NDCG@5	&	0.0236 	&	0.0263 	&	0.0342 	&	0.0352 	&	0.0352 	&	0.0354 	&$\underline{	0.0361 	}$&$\mathbf{	0.0377 	}\pm	0.0008 	$&$	4.43 	\%$\\
&	NDCG@10	&	0.0281 	&	0.0317 	&	0.0428 	&	0.0446 	&	0.0443 	&	0.0442 	&$\underline{	0.0454 	}$&$\mathbf{	0.0476 	}\pm	0.0015 	$&$	4.84 	\%$\\\midrule\multirow{4}{*}{Sports}
&	HR@5	&	0.0218 	&	0.0227 	&	0.0311 	&	0.0328 	&	0.0326 	&	0.0328 	&$\underline{	0.0334 	}$&$\mathbf{	0.0354 	}\pm	0.0015 	$&$	5.99 	\%$\\
&	HR@10	&	0.0336 	&	0.0374 	&	0.0479 	&	0.0501 	&	0.0498 	&	0.0502 	&$\underline{	0.0506 	}$&$\mathbf{	0.0528 	}\pm	0.0014 	$&$	4.35 	\%$\\
&	NDCG@5	&	0.0127 	&	0.0149 	&	0.0189 	&	0.0204 	&	0.0208 	&	0.0206 	&$\underline{	0.0214 	}$&$\mathbf{	0.0227 	}\pm	0.0006 	$&$	6.07 	\%$\\
&	NDCG@10	&	0.0169 	&	0.0194 	&	0.0260 	&	0.0260 	&	0.0262 	&	0.0264 	&$\underline{	0.0271 	}$&$\mathbf{	0.0287 	}\pm	0.0015 	$&$	5.90 	\%$\\\midrule\multirow{4}{*}{ML-1M}
&	HR@5	&	0.1087 	&	0.1147 	&	0.1987 	&	0.2041 	&	0.2038 	&	0.2039 	&$\underline{	0.2043 	}$&$\mathbf{	0.2113 	}\pm	0.0017 	$&$	3.43 	\%$\\
&	HR@10	&	0.1904 	&	0.1975 	&	0.2904 	&	0.2933 	&	0.2946 	&	0.2944 	&$\underline{	0.2951 	}$&$\mathbf{	0.3045 	}\pm	0.0021 	$&$	3.19 	\%$\\
&	NDCG@5	&	0.0638 	&	0.0662 	&	0.1346 	&	0.1389 	&	0.1390 	&	0.1392 	&$\underline{	0.1402 	}$&$\mathbf{	0.1469 	}\pm	0.0013 	$&$	4.78 	\%$\\
&	NDCG@10	&	0.0910 	&	0.0928 	&	0.1634 	&	0.1683 	&	0.1680 	&	0.1682	&$\underline{	0.1696 	}$&$\mathbf{	0.1763 	}\pm	0.0010 	$&$	3.95 	\%$\\\midrule\multirow{4}{*}{ML-20M}
&	HR@5	&	0.1143 	&	0.1287 	&	0.2079 	&	0.2102 	&	0.2098 	&	0.2096 	&$\underline{	0.2113 	}$&$\mathbf{	0.2193 	}\pm	0.0014 	$&$	3.79 	\%$\\
&	HR@10	&	0.2152 	&	0.2271 	&	0.2802 	&	0.2995 	&	0.3001 	&	0.3007 	&$\underline{	0.3011 	}$&$\mathbf{	0.3113 	}\pm	0.0018 	$&$	3.39 	\%$\\
&	NDCG@5	&	0.0717 	&	0.0891 	&	0.1399 	&	0.1421 	&	0.1428 	&	0.1433 	&$\underline{	0.1436 	}$&$\mathbf{	0.1503 	}\pm	0.0012 	$&$	4.69 	\%$\\
&	NDCG@10	&	0.1013 	&	0.1131 	&	0.1652 	&	0.1726 	&	0.1735 	&	0.1739 	&$\underline{	0.1741 	}$&$\mathbf{	0.1805 	}\pm	0.0009 	$&$	3.68 	\%$\\\midrule\multirow{4}{*}{Yelp}
&	HR@5	&	0.0155 	&	0.0233 	&	0.0433 	&	0.0454 	&	0.0429 	&	0.0439 	&$\underline{	0.0461 	}$&$\mathbf{	0.0481 	}\pm	0.0004 	$&$	4.34 	\%$\\
&	HR@10	&	0.0268 	&	0.0342 	&	0.0617 	&	0.0647 	&	0.0614 	&	0.0618 	&$\underline{	0.0651 	}$&$\mathbf{	0.0675 	}\pm	0.0011 	$&$	3.69 	\%$\\
&	NDCG@5	&	0.0103 	&	0.0122 	&	0.0307 	&	0.0332 	&	0.0324 	&	0.0333 	&$\underline{	0.0335 	}$&$\mathbf{	0.0358 	}\pm	0.0012 	$&$	6.87 	\%$\\
&	NDCG@10	&	0.0133 	&	0.0151 	&	0.0356 	&	0.0394 	&	0.0383 	&	0.0388 	&$\underline{	0.0399 	}$&$\mathbf{	0.0418 	}\pm	0.0004 	$&$	4.76 	\%$\\\midrule\multirow{4}{*}{Life Service}
&	HR@5	&	0.0108 	&	0.0117 	&	0.0179 	&	0.0198 	&	0.0193 	&	0.0193 	&$\underline{	0.0204 	}$&$\mathbf{	0.0216 	}\pm	0.0013 	$&$	5.88 	\%$\\
&	HR@10	&	0.0142 	&	0.0161 	&	0.0225 	&	0.0249 	&	0.0246 	&	0.0248 	&$\underline{	0.0252 	}$&$\mathbf{	0.0268 	}\pm	0.0012 	$&$	6.35 	\%$\\
&	NDCG@5	&	0.0066 	&	0.0077 	&	0.0118 	&	0.0142 	&	0.0145 	&	0.0146 	&$\underline{	0.0149 	}$&$\mathbf{	0.0164 	}\pm	0.0013 	$&$	6.71 	\%$\\
&	NDCG@10	&	0.0096 	&	0.0112 	&	0.0158 	&	0.0187 	&	0.0189 	&	0.0185 	&$\underline{	0.0192 	}$&$\mathbf{	0.0205 	}\pm	0.0008 	$&$	6.77 	\%$\\
\bottomrule
\end{tabular}
    \label{tab:result}
\end{table*}

\begin{table}[htpb]
    \centering
\caption{Performance improvement on the state-of-the-art SR model FMLP.}   
\begin{tabular}{c|c|cc|c}
\toprule 
Dataset  &	Metric	&	FMLP	& FMLP+HPM &	FMLP+RCL	\\
\midrule\multirow{4}{*}{ML-20M} 
&	HR@5	&	0.1384 	&	0.2154 	&$	0.2233 	\pm	0.0007	$	\\
&	HR@10	&	0.2398 	&	0.3062 	&$	0.3144 	\pm	0.0011	$	\\
&	NDCG@5	&	0.0925 	&	0.1477 	&$	0.1528 	\pm	0.0013	$	\\
&	NDCG@10	&	0.1283 	&	0.1783 	&$	0.1826 	\pm	0.0007	$	\\\midrule\multirow{4}{*}{Yelp}
&	HR@5	&	0.0197 	&	0.0473 	&$	0.0496 	\pm	0.0012	$	\\
&	HR@10	&	0.0321 	&	0.0657 	&$	0.0686 	\pm	0.0011	$	\\
&	NDCG@5	&	0.0148 	&	0.0357 	&$	0.0379 	\pm	0.0014	$	\\
&	NDCG@10	&	0.0185 	&	0.0423 	&$	0.0452 	\pm	0.0008	$	\\
\bottomrule
\end{tabular}
    \label{tab:newbase}
\end{table}
\subsubsection{Baselines} 
We compare our proposed RCL with various existing baselines. We consider three categories (I: base model without CL, II: augmentation-based method, III: SCL-based method) of comparison methods as follows.
\begin{itemize}[leftmargin=10pt, rightmargin=0pt]
    \item \textbf{SASRec~(I)}~\cite{SASRec} is a single-directional self-attention model. It is a strong baseline in the sequential recommendation.
    \item \textbf{CL4SRec~(II)}~\cite{CL4SRec} uses item cropping, masking, and reordering as augmentations for contrastive learning. It is the first contrastive learning method for sequential recommendation.
    \item \textbf{CT4Rec~(II)}~\cite{CT4Rec} offers a model augmentation-based contrastive learning approach for sequential recommendation, which adds two extra training objectives that ensure consistency in user representations across different dropout masks during training.
    \item \textbf{MCLRec~(II)}~\cite{MCLRec} is an augmentation-based method for sequential recommendation, which contrasts data/model-level augmented views for adaptively capturing the informative features hidden in stochastic data augmentation.
    \item \textbf{DuoRec~(III)}~\cite{DuoRec} \nnhl{utilizes both sequences with same target item and model-level augmented sequences as positive samples for contrastive learning.}
    \item \textbf{ContraRec~(III)}~\cite{ContraRec} introduces a supervised contrastive learning method for sequential recommendation by utilizing sequences with same target item as positive samples.
    \item \textbf{HPM~(III)}~\cite{HPM} is a unified framework for sequential recommendation that leverages contrastive learning to optimize hierarchical self-supervised signals in user interaction sequences, accommodating various base sequence encoders.
\end{itemize}
\subsubsection{Implementation} 
RCL and all baselines use SASRec as the base model. The embedding size and hidden size are set to 64. The numbers of layers and heads in the Transformer are set to 2. The Dropout~\cite{DuoRec} rate on the embedding matrix and the Transformer module is chosen from $\{0.1,0.2,0.3,0.4,0.5\}$. The training batch size is set to 256. We use the Adam~\cite{SASRec} optimizer with the learning 0.001. $\lambda$ in Equation~(\ref{eq:total}) is chosen from $\{0.1,0.2,0.3,0.4,0.5\}$. $\alpha$ in Section~\ref{sec:weak} is chosen from $\{0.025,0.05,0.1,0.15,0.2,0.25\}$. 
Temperature $\tau$ in Equation~(\ref{eq:infonce}) is chosen from $\{0.01,0.05,0.1,0.5,1,5\}$.

\subsection{RQ1: Overall Performance}

In this study, we assess the overall performance of RCL against various baselines, with the comparative results delineated in Table~\ref{tab:result}. For this comparison, we utilize the 2-gram similarity variant of RCL. \newhl{From the collected data, we make several observations. First, the augmentation-based methods—CL4SRec, MCLRec, and CT4Rec—consistently outperform base model that do not utilize contrastive learning, highlighting the significance of augmentation-based contrastive learning in enhancing sequence representations. Second, among SCL-based methods, HPM outshines other augmentation-based methods. Third, RCL achieves 4.36\%, 5.57\%, 3.83\%, 3.88\%, 4.91\%, and 6.42\% improvements over state-of-the-art methods across six datasets, evidencing the effectiveness of the proposed dual-tiered positive sample selection module and relative contrastive learning module. Section~\ref{timecost} introduces a way to use the GPU for similarity calculating acceleration. By setting the thread number of the GPU as 256 and the similar sequence ratio $\alpha$ as 0.05,
GPU-based sequence similarity calculation can reduce 99.0\% time cost and 95.1\% memory cost \nnhl{on two largest public datasets, \textbf{ML-20M} and \textbf{Yelp},} in average compared to the CPU-based method.}

We further conduct a supplementary performance evaluation as shown in Table~\ref{tab:newbase}, examining RCL's impact on a more advanced base model called FMLP. FMLP~\cite{fmlp} is an all-MLP model utilizing a learnable filter-enhanced block for noise reduction in the embedding matrix. We report the results on the two largest public datasets, ML-20M and Yelp, and observe similar improvements on other four datasets. Table~\ref{tab:newbase} shows that FMLP combined with RCL considerably improves performance by 3.97\% and 5.57\% on the two datasets against the state-of-the-art SCL method~(HPM), respectively. These findings suggest that RCL can consistently boost the capabilities of various base models, including both transformer-based and MLP-based models.



\subsection{RQ2: Ablation Study}
\label{sec:ablation}
\begin{figure}[htpb]
    \centering
	\begin{minipage}[b]{\linewidth}
		\begin{subfigure}
   \centering
	\includegraphics[width=0.95\textwidth]{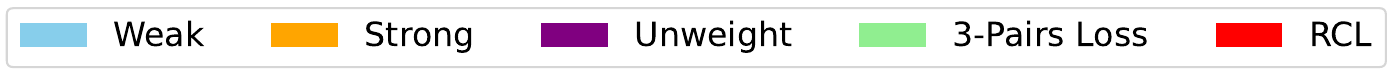}
	\end{subfigure}
 \end{minipage}
 \setcounter{subfigure}{0}
 \centering
 \begin{minipage}[b]{\linewidth}
  \subfigure[ML-20M-HR@5]{
    \includegraphics[width=0.49\linewidth]{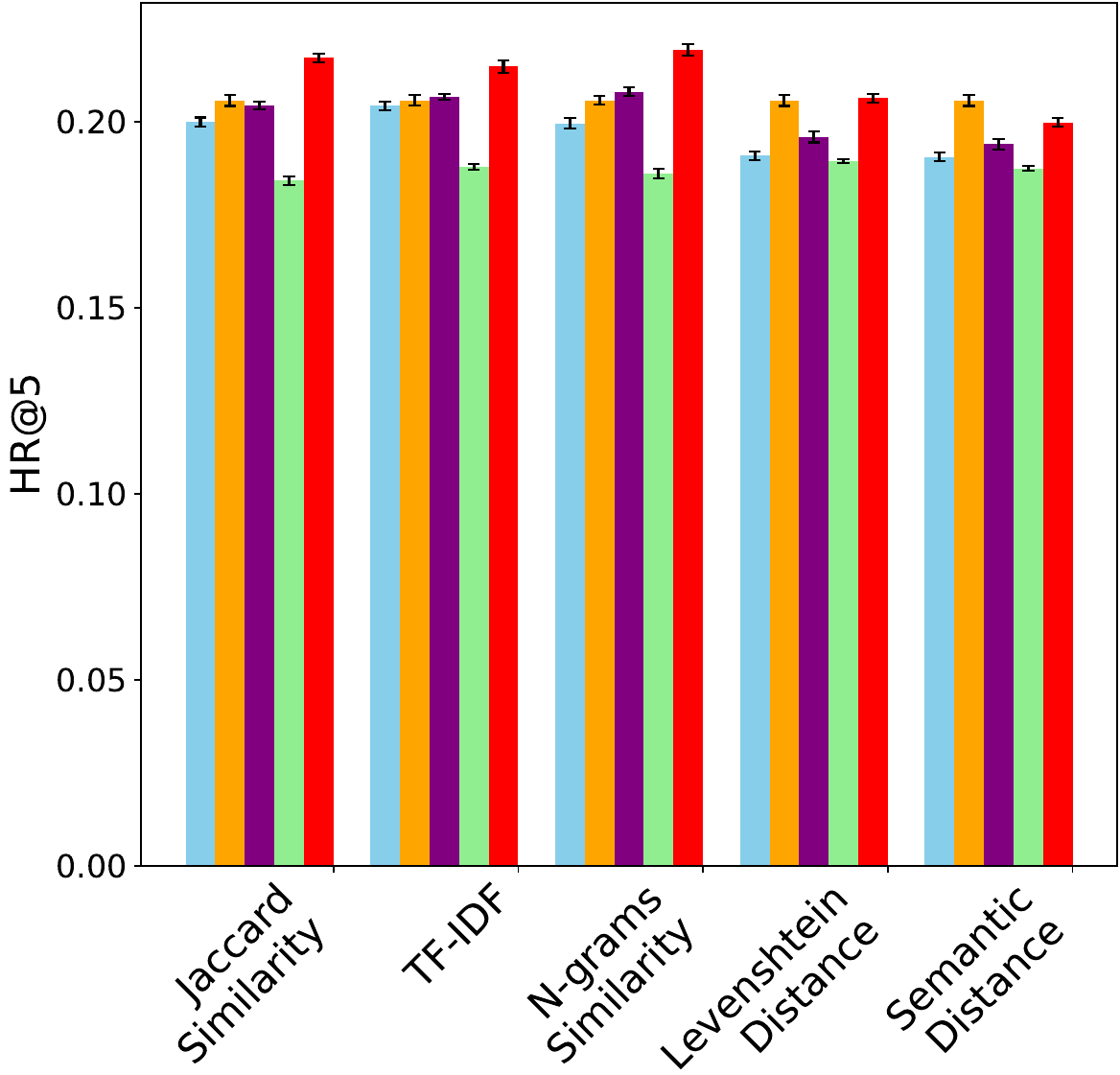}
    \label{fig:beautyhr}
    }
  \subfigure[ML-20M-NDCG@5]{
    \includegraphics[width=0.49\linewidth]{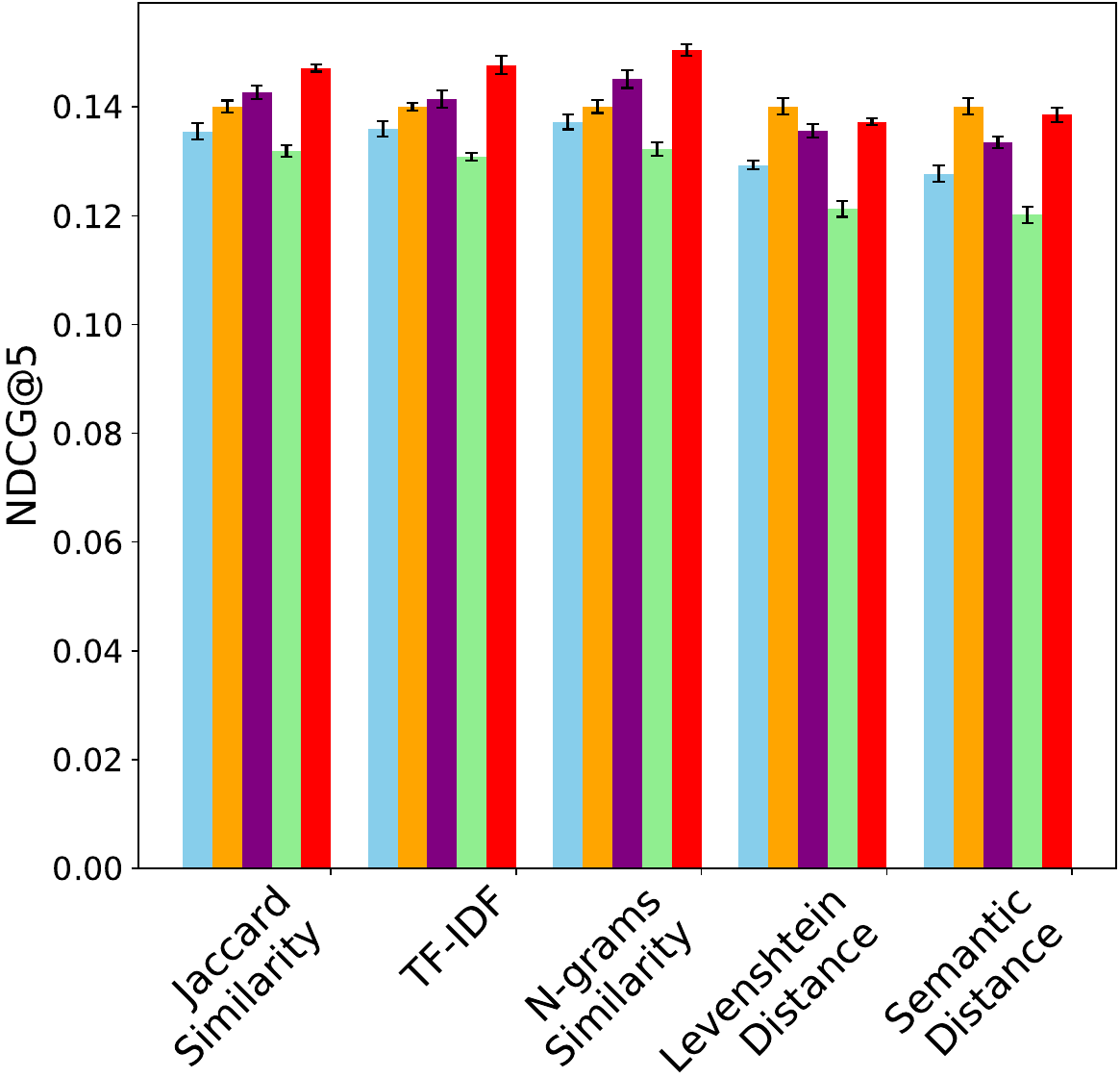}
    \label{fig:beautyndcg}
}
\end{minipage}
\centering
\begin{minipage}[b]{\linewidth}
  \subfigure[Yelp-HR@5]{
    \includegraphics[width=0.49\linewidth]{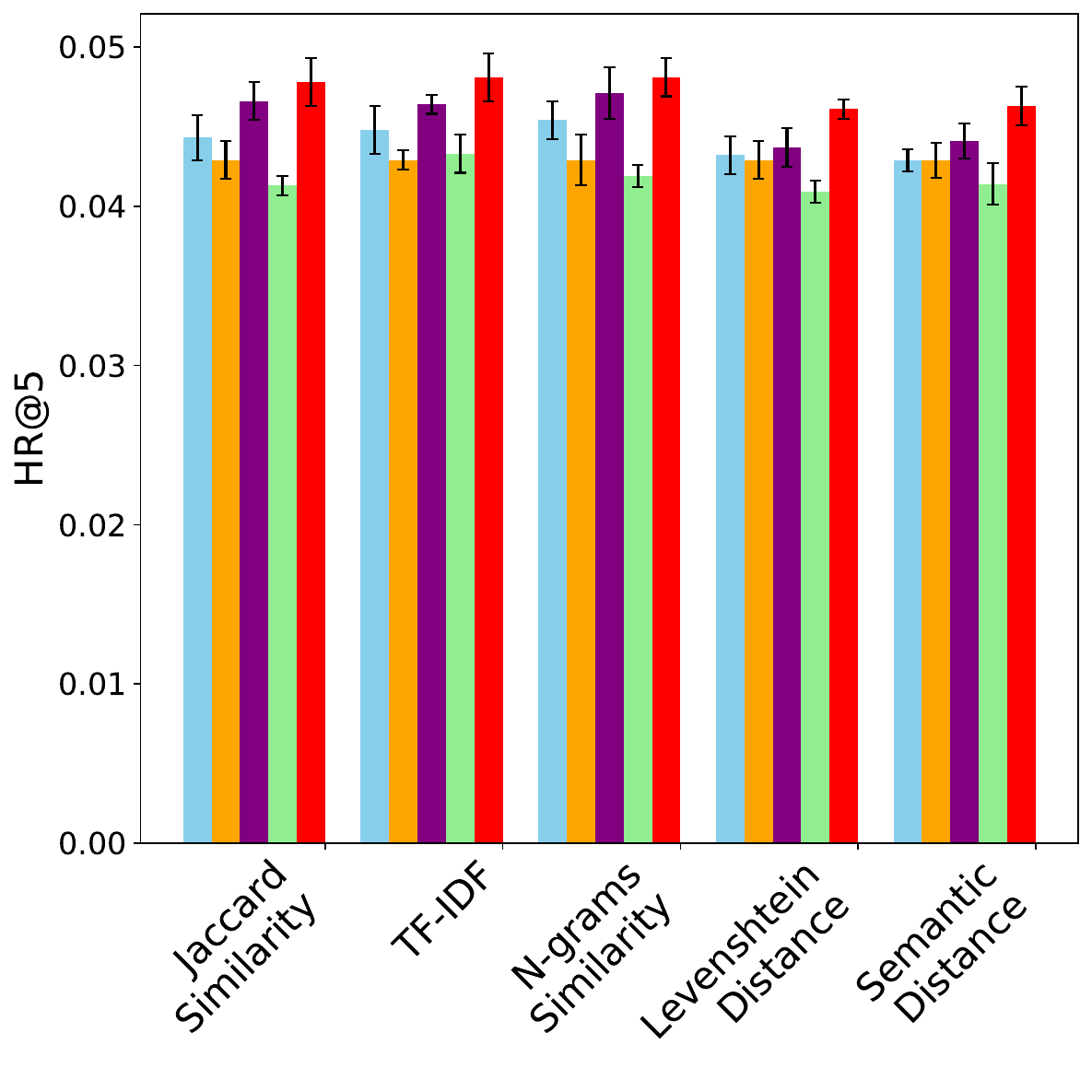}
    \label{fig:yelphr}
  }
  \subfigure[Yelp-NDCG@5]{
    \includegraphics[width=0.49\linewidth]{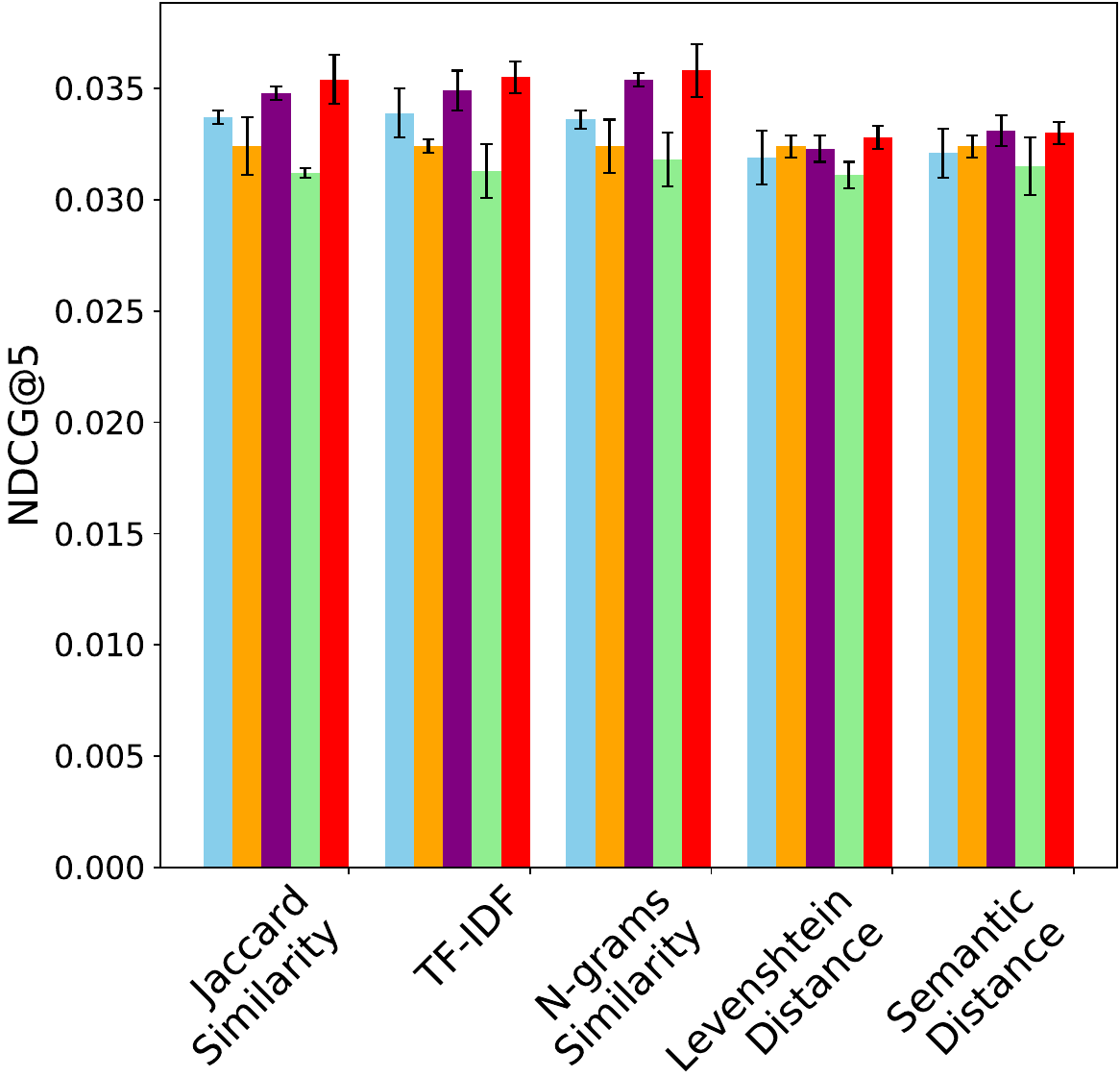}
    \label{fig:yelpndcg}
  }
  \end{minipage}
  \caption{Ablation study of different similarity metrics and loss functions.}
  \label{fig:ablation}
\end{figure}

In this section, we conduct an ablation study on two largest public datasets, \textbf{ML-20M} and \textbf{Yelp}, to assess the impact of different similarity metrics and loss functions in RCL on recommendation performance. Similar observations are made across the other four datasets and HR/NDCG@10. For similarity metrics (mentioned in Section~\ref{sec:weak}), five metrics are evaluated. For loss functions, the following variants are considered (from Section~\ref{sec:loss}):

\begin{itemize}[leftmargin=10pt, rightmargin=0pt]
    \item \textbf{Weak} utilizes only weak positive samples using $\mathcal{L}^{RCL}$ from Eq.~(\ref{eq:weak}).
    \item \textbf{Strong} solely relies on strong positive samples with $\mathcal{L}^{pair}$ from Eq.~(\ref{eq:infonce}). In cases where same-target sequences are absent, augmented user representation serves as positive samples instead~\cite{DuoRec}.
    \item \textbf{Unweight} represents the unweighted RCL version using $\mathcal{L}^{RCL}$ in Eq.~(\ref{eq:unweight}).
    \item \textbf{RCL} incorporates the proposed weighted RCL version using $\mathcal{L}^{wRCL}$ in Eq.~(\ref{eq:RCL}).
    \item \textbf{3-Pairs Loss} combines three infoNCE losses:
    \begin{equation}
        \mathcal{L}^{3-Pairs} = \sum_{u \in \mathcal{U}} \left(\mathcal{L}^{\text{S-W}}_u+\mathcal{L}^{\text{pair}}(S_u, S_{u,a})+\mathcal{L}^{\text{pair}}(S_u, S_{u,b})\right).
        \label{eq:sw}
    \end{equation}
    For sequence $S_u$, $\mathcal{L}^{\text{S-W}}_u$ treats same-target/similar sequences as positive/negative samples to encourage same-target sequences to have closer representations with $S_u$ than similar sequences. $\mathcal{L}^{\text{pair}}(S_u, S_{u,a})$/$\mathcal{L}^{\text{pair}}(S_u, S_{u,b})$ treat same-target/similar sequences as positive samples and treat the other sequences except $S_u$ in the minibatch as negative samples.
\end{itemize}


Figure~\ref{fig:ablation} shows the performance of RCL using different similarity metrics and loss functions. We summarize four observations. \myhl{First, both \textbf{Unweight} and \textbf{RCL} outperform \textbf{Weak} and \textbf{Strong}, indicating the effectiveness of employing both strong and weak positive samples. Using only similar sequences might introduce more noise, while relying solely on same-target sequences could introduce noise too if they're unavailable and change to augmented representation instead.}
Second, \textbf{RCL} performs better than \textbf{Unweight} on both datasets, highlighting the effectiveness of the similarity-based reweighting mechanism (Section~\ref{sec:loss}). 
Third, \textbf{3-Pairs Loss} significantly underperforms the other four groups, probably 9 due to $\mathcal{L}^{\text{S-W}}$ in Eq.~(\ref{eq:sw}) pushing weak positive samples further from the center sequence $S_u$ than negative samples. 
\myhl{Fourth, the order-sensitive groups~(Levenshtein Distance and Semantic Distance) underperform order-insensitive ones, probably because order-sensitive metrics might be overly strict by filtering sequences with many co-appearing items but differing interaction orders. However, \textit{2-grams Similarity} slightly outperforms \textit{Jaccard Similarity} and \textit{TF-IDF}, suggesting the importance of considering local order.}


\subsection{RQ3: Parameter Sensitivity}
    
   
\begin{figure}[t]
    \centering
	\begin{minipage}[b]{\linewidth}
 \centering
		\begin{subfigure}
   \centering
	\includegraphics[width=0.6\textwidth]{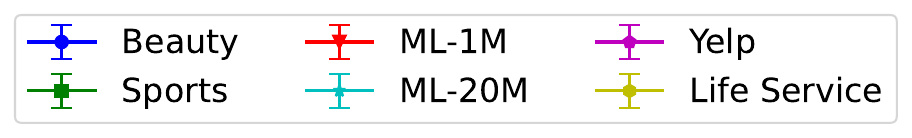}
	\end{subfigure}
 \end{minipage}
 \setcounter{subfigure}{0}
\begin{minipage}[b]{\linewidth}
  \subfigure[$\alpha$-HR@5]{
    \includegraphics[width=0.49\linewidth]{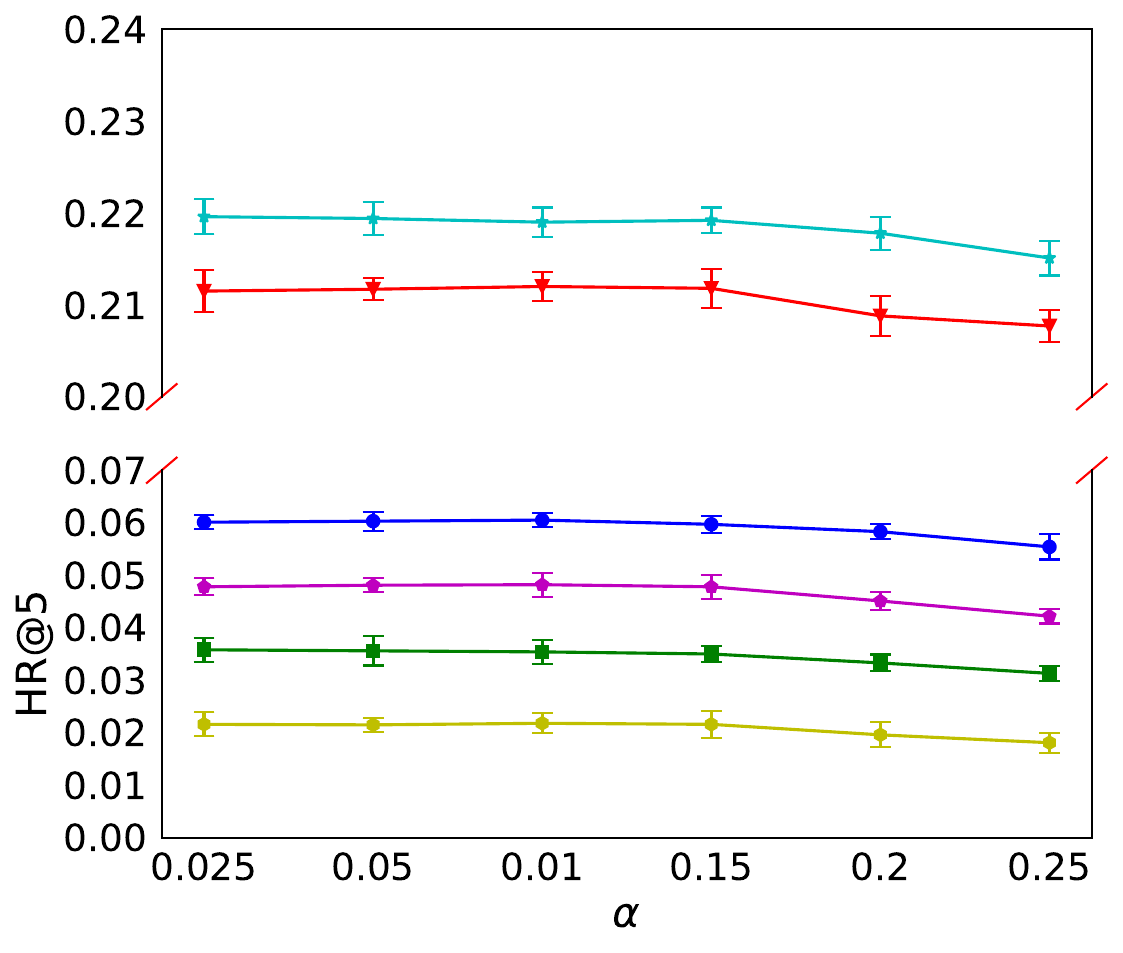}
    \label{fig:alphahr}
  }
  \subfigure[$\alpha$-NDCG@5]{
    \includegraphics[width=0.49\linewidth]{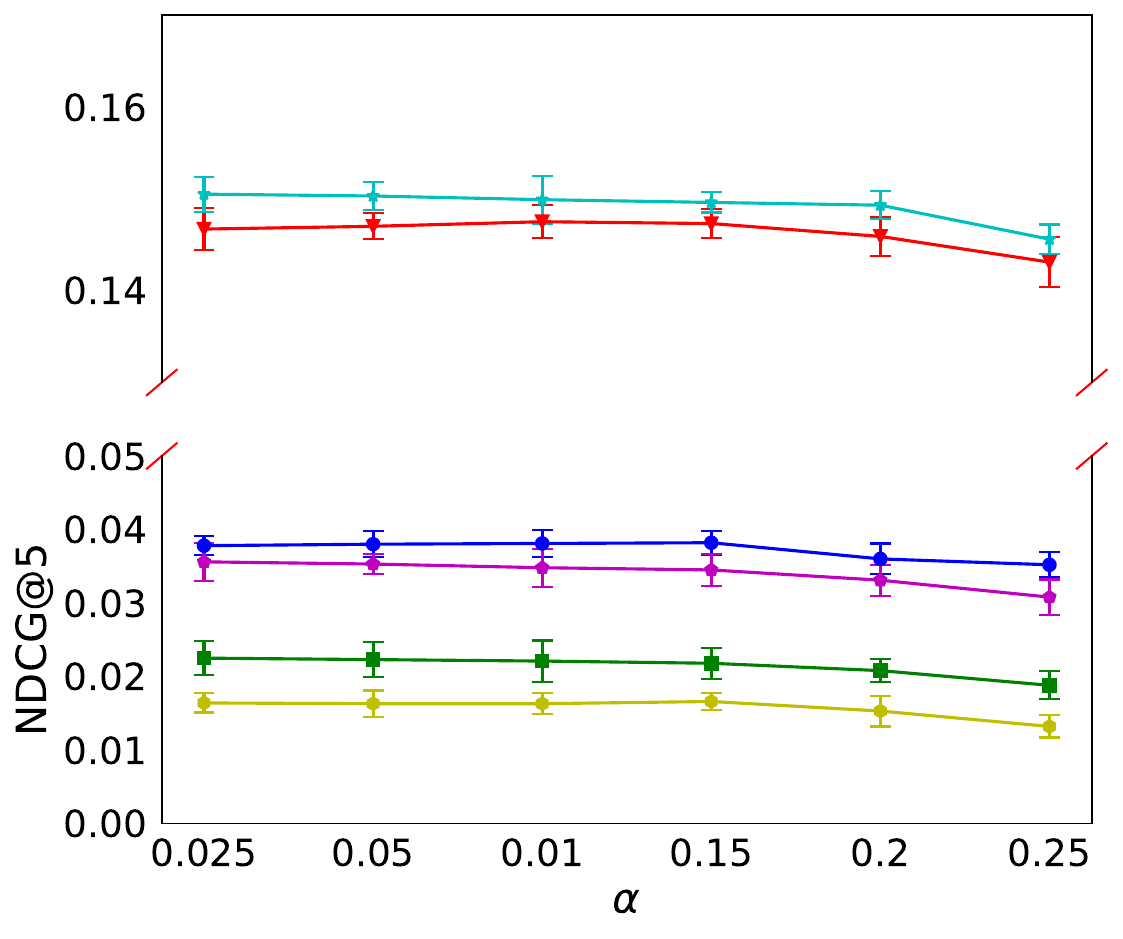}
    \label{fig:alphandcg}
  }
  \end{minipage}
  \caption{Performance with different top similar sequence ratios $\alpha\%$ on \textbf{Beauty} and \textbf{Yelp} datasets.}
  \label{fig:parameter}
\end{figure}

We provide the sensitivity results of top similar sequence ratio $\alpha\%$ on RCL performance in Figure~\ref{fig:parameter}. We focus on the \textbf{Beauty} and \textbf{Yelp} datasets, evaluating on HR/NDCG@5. Similar observations hold for other datasets and evaluation metrics. The top similar sequence ratio $\alpha$ in Section~\ref{sec:weak}, chosen from $\{0.025, 0.05, 0.1, 0.15, 0.02, 0.025\}$, displays consistent performance with smaller values but exhibits roughly 10\% performance drop with larger ratios. As $\alpha$ increases, sequences with low similarity are more likely to be sampled as weak positive samples. The performance drop might be attributed to such noisy weak positive samples.

\subsection{RQ4: Case Study}
\begin{figure}[htpb]
    \centering
	\begin{minipage}[b]{\linewidth}
    \centering
		\begin{subfigure}
   \centering
	\includegraphics[width=0.99\textwidth]{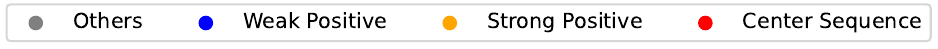}
	\end{subfigure}
 \end{minipage}
 \setcounter{subfigure}{0}
 \centering
 \begin{minipage}[b]{\linewidth}
 \centering
  \subfigure[SASRec-Beauty]{
    \includegraphics[width=0.30\linewidth]{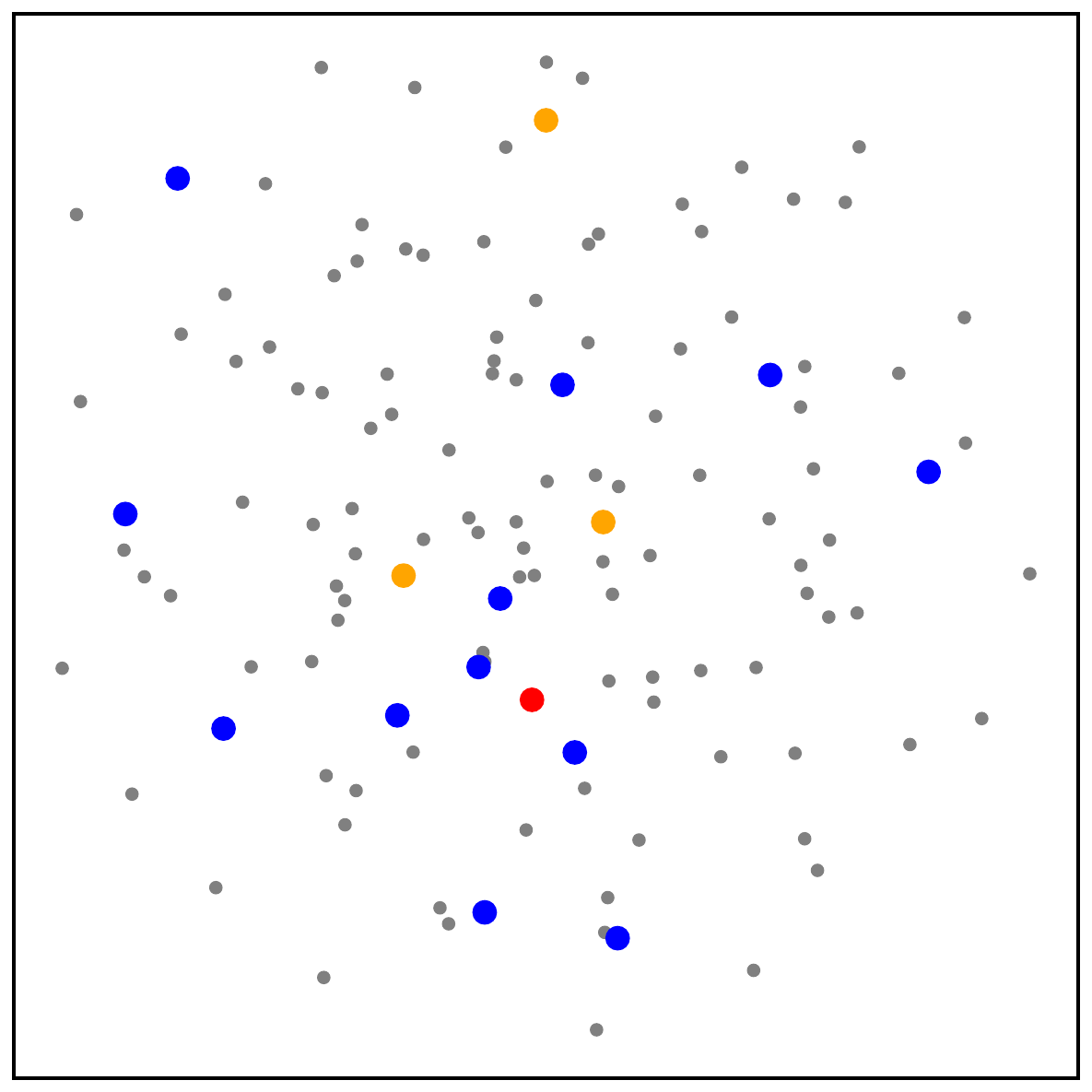}
    \label{fig:case_sasrec}
    }
  \subfigure[HPM-Beauty]{
    \includegraphics[width=0.30\linewidth]{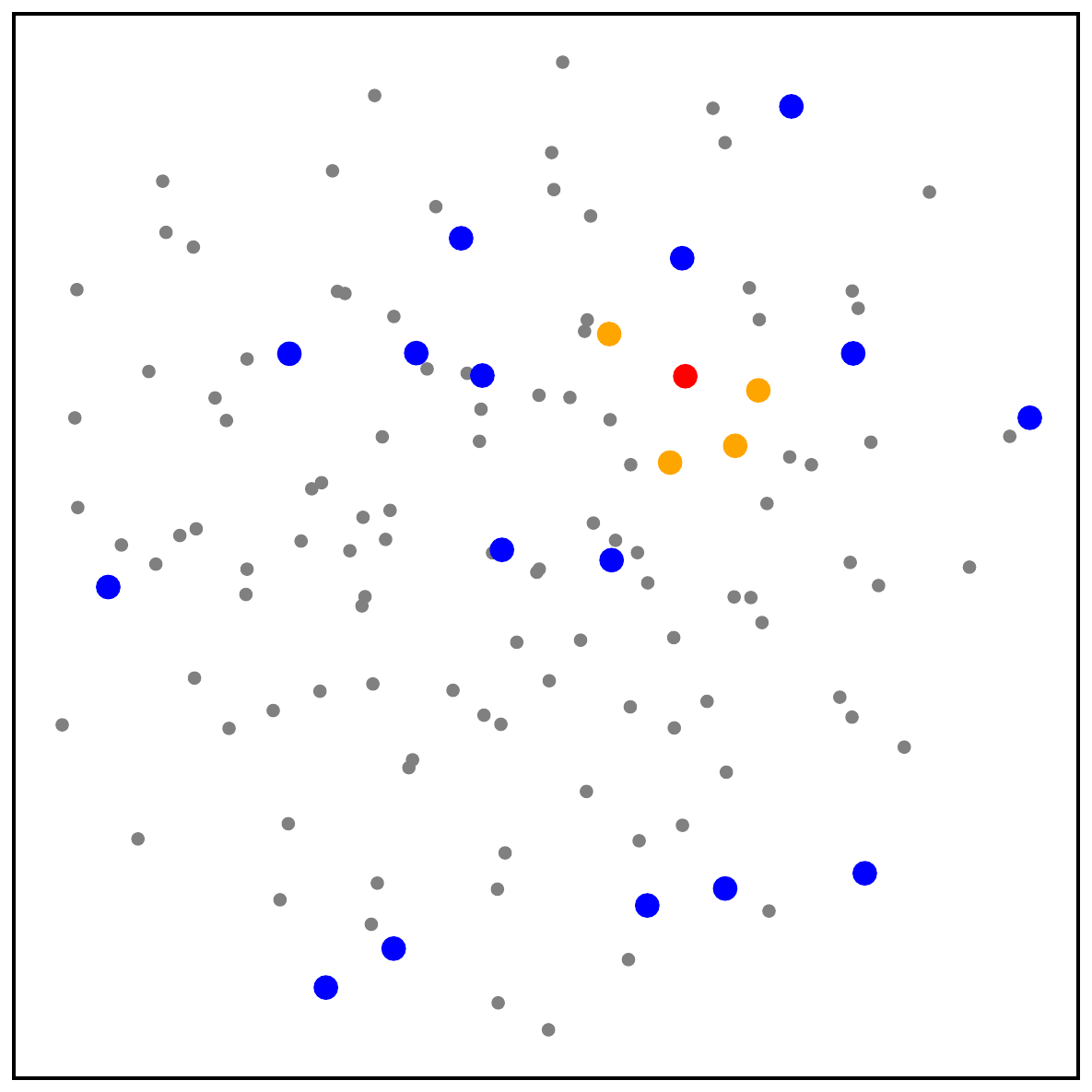}
    \label{fig:case_scl}
    }
\subfigure[RCL-Beauty]{
      \includegraphics[width=0.30\linewidth]{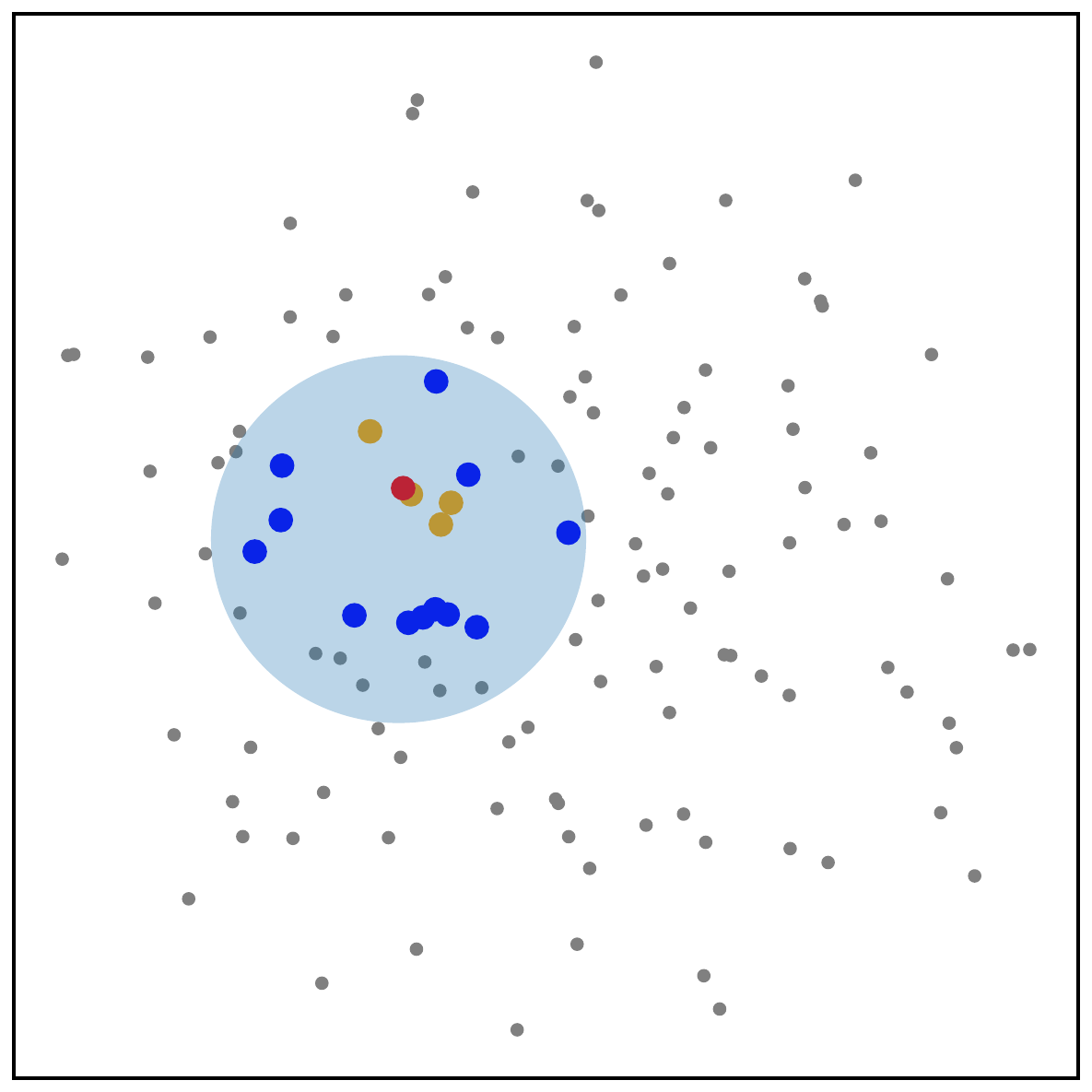}
      \label{fig:case_rcl}
    }
\end{minipage}
\centering
\begin{minipage}[b]{\linewidth}
  \centering
   \subfigure[SASRec-Yelp]{
     \includegraphics[width=0.30\linewidth]{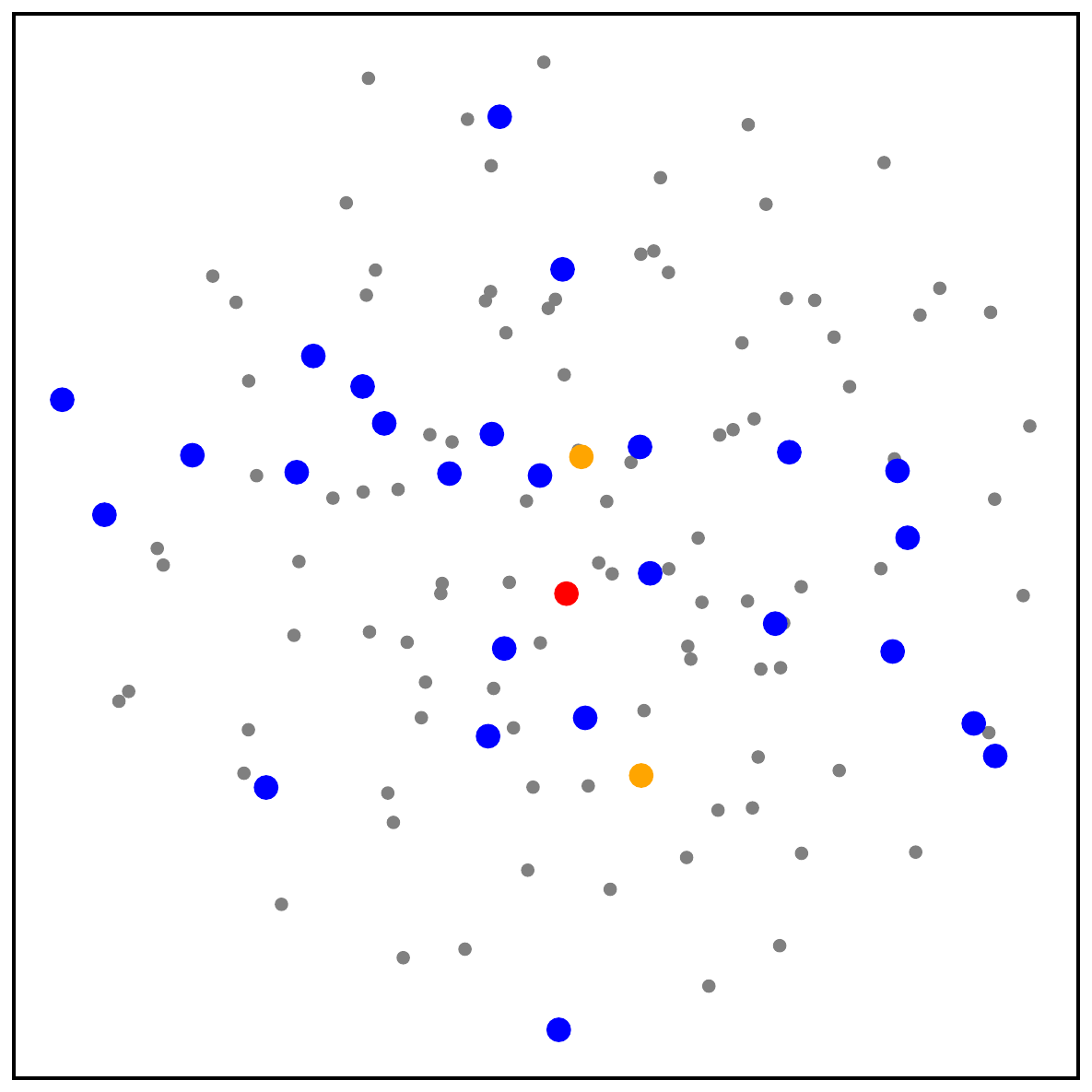}
     \label{fig:case_sasrec2}
     }
   \subfigure[HPM-Yelp]{
     \includegraphics[width=0.30\linewidth]{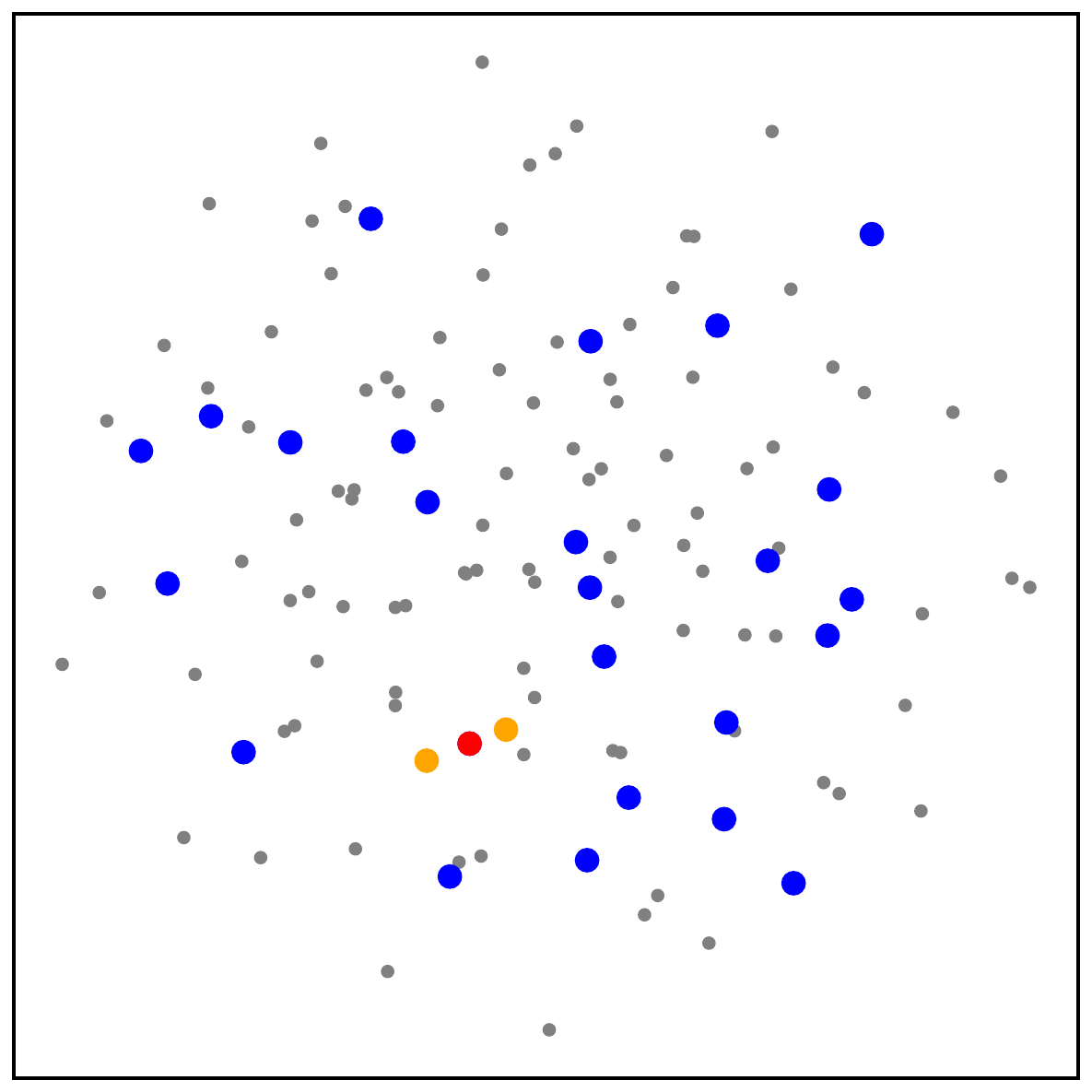}
     \label{fig:case_scl2}
     }
 \subfigure[RCL-Yelp]{
       \includegraphics[width=0.30\linewidth]{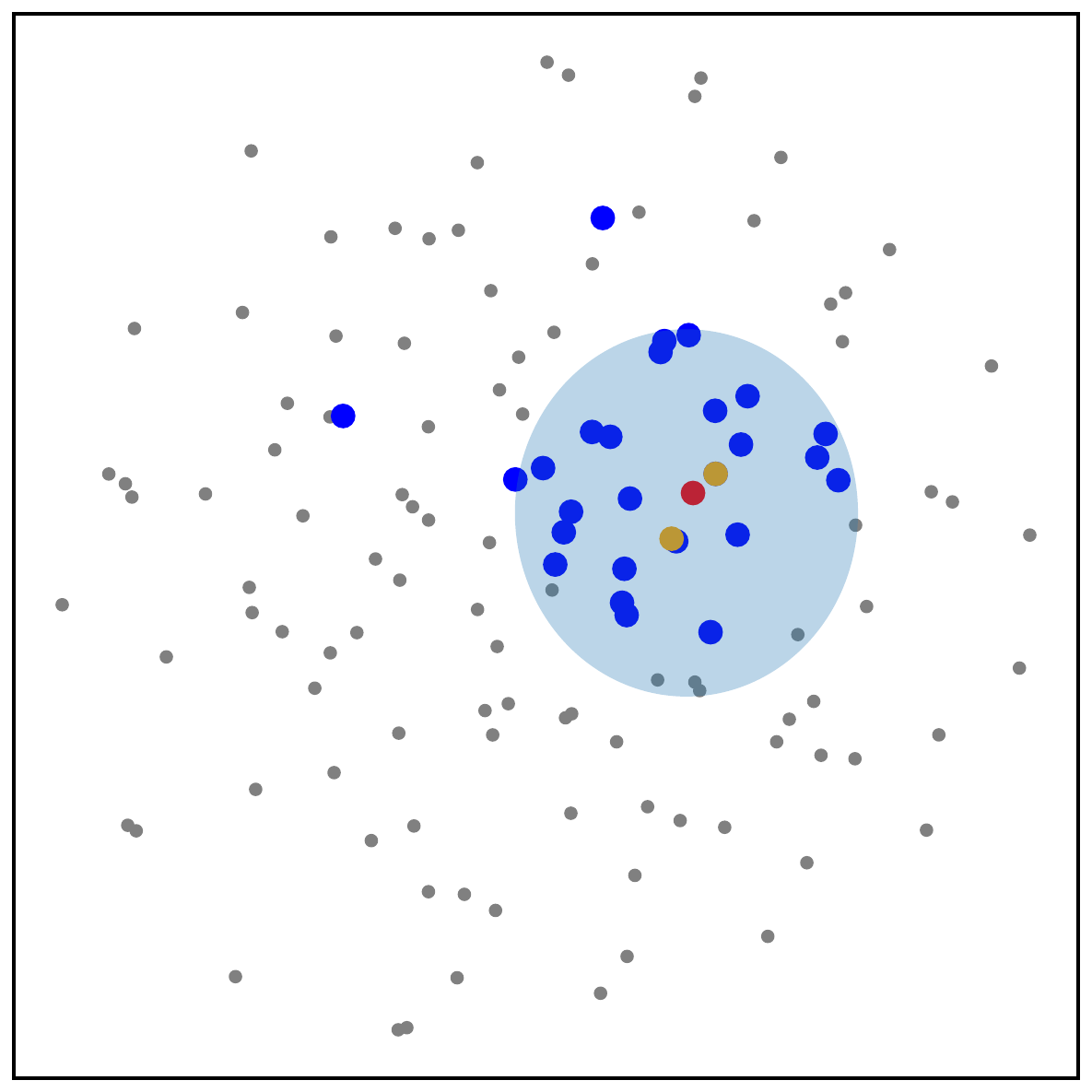}
       \label{fig:case_rcl2}
     }
 \end{minipage}
  \caption{t-SNE results of two sequences from \textbf{Beauty} and \textbf{Yelp} for RCL and the compared baselines.} 
  \label{fig:casestudy}
\end{figure}

\begin{figure}[htpb]
  \centering
     \includegraphics[width=0.99\linewidth]{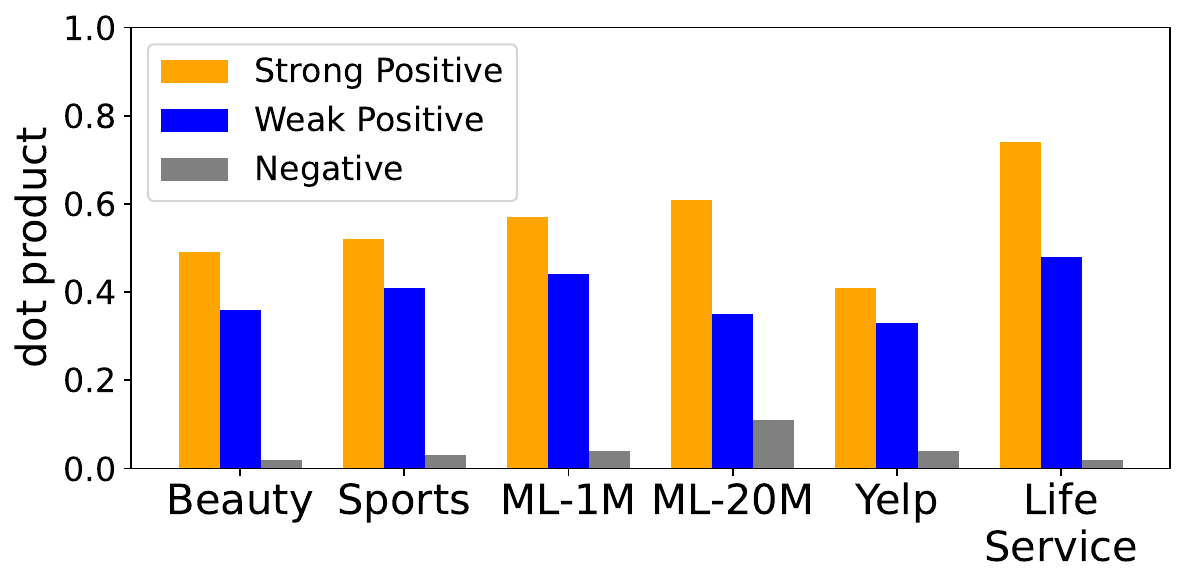}
     \label{fig:case_dots}
  \caption{The average dot products of the center sequence.}
  \label{fig:case_dots}
\end{figure}

\myhl{In this section, we conduct a case study by acquiring the t-SNE embeddings of two center sequences along with their associated strong positive samples, weak positive samples, and negative samples for RCL. For brevity, we randomly sample 128 sequences for each center sequence. For comparison, we also provide the t-SNE embeddings of these 128 sequences in SASRec and HPM~(a representative SCL method).
We observe that in HPM and RCL, the strong positive samples are closest to the center sequence, indicating the infoNCE loss denoted in Eq.~(\ref{eq:infonce}) is effective. With RCL, the weak positive samples are situated closer to the center sequence compared to negative samples but not as close as strong positive samples, which demonstrates the effectiveness of the restraint mechanism in Eq.~(\ref{eq:RCL}). Figure~\ref{fig:case_dots} provides the average dot products of the center sequences with their associated strong positive samples, weak positive samples, and negative samples on six datasets respectively, further underlining the efficacy of the RCL loss function.}

\begin{table}[t]
    \centering
    \caption{The improvements of RCL during online evaluation on the Average View Time per User (AVTU) metric.}
    \begin{tabular}{c|ccccccc}
        \toprule
         Day& 1 & 2 & 3 & 4 & 5 & 6 & 7\\
         \midrule
         Improvement(\%) & 0.32 & 0.36 & 0.93 & 0.93 & 1.03 & 1.27 & 0.20 \\
         \bottomrule
    \end{tabular}
    
    \label{tab:online}
\end{table}

\subsection{RQ5: A/B Testing}
\newhl{We further evaluate RCL on a private online recommendation platform which uses MIND~\cite{MIND} as the base sequential recommendation model. We use RCL for 20\% users and use SCL~(only strong positive samples) for the others and compare the performance between RCL and SCL. Since the online platform produces 20+ million interaction sequences every day, we only calculate similarities of sequences whose target items have the same category to guarantee the sampling efficiency. The evaluation time span is 2024/06/07-2024/06/13. This platform uses Average View Time per User~(AVTU) as the main evaluation metric which indicates user engagement and the overall user experience with the platform. Table~\ref{tab:online} shows the improvements of RCL during 7-day recommendation, which testifies the effectiveness of RCL for real recommendation systems.} 


\section{Related Work} \label{sec:related-work}
\subsection{Sequential Recommendation} 
Existing sequential recommendation models~\cite{time_lstm,din,fpmc,hrm,IMSR,FeSAIL}
mainly rely on recurrent neural networks like GRU \cite{GRU4Rec} or attention mechanisms such as transformer ~\cite{SASRec, BERT4Rec, S3Rec, ema} as sequence encoders. Some works ~\cite{FGNN, GAG, PosRec} also incorporate graph neural networks to model sequences. 
Most models ~\cite{din,SASRec} center on the next-item prediction task, which is inherently suitable for predicting subsequent items. 
The supervision of next-item prediction is limited to data sparsity issue ~\cite{MMInfoRec} and auxiliary tasks such as masked prediction have a semantic gap to the recommendation task. 

\subsection{Contrastive Learning} 
Contrastive learning (CL) has found extensive applications~\cite{CPC, DIM, MoCo, SimCLR, SimSiam} in deep learning to learn latent representations. Based on whether labels are used when generating positive samples, CL can be grouped into self-supervised CL and supervised CL. 
Some works~\cite{S3Rec, CL4SRec, DHCN, MHCN, MMInfoRec} apply self-supervised CL for sequential recommendation. For example, 
\newhl{CoSeRec~\cite{CoSeRec} and TiCoSeRec~\cite{TiCoSeRec} perform item-level CL based on model-based augmentations.} Recently, some works\cite{ContraRec,DuoRec,HPM} propose to apply supervised contrastive learning by taking the sequences with the same target item as positive samples and maximize their similarity. 
Different from existing methods, we focus on using similar sequences as supplements to positive samples and provide extra training signals for optimizing sequential recommendation models.

\section{Conclusion}
\label{sec:conclusion}
In this work, we introduce a novel framework called \textbf{R}elative \textbf{C}ontrastive \textbf{L}earning (RCL) for sequential recommendation, which treats similar sequences as additional positive samples. Our approach comprises a dual-tiered positive sample selection module, and a relative contrastive learning module. The former module utilizes same-target sequences as strong positive samples and treats similar sequences as weak positive samples. 
The latter module utilizes a weighted relative contrastive loss function 
to draw the representations of strong positive samples relatively closer to center sequence. 
We apply RCL to two important sequential recommendation models, and our empirical results reveal that RCL averagely achieves 4.88\% improvement against the state-of-the-art methods across five public datasets and one private dataset.

\section*{Acknowledgements}
\newhl{
Yanyan Shen is the corresponding author and is partially supported by the National Key Research and Development Program of China (2022YFE0200500) and the Shanghai Municipal Science and Technology Major Project (2021SHZDZX0102).
\nnhl{We would like to extend our special thanks to Peng Yan and Dongbo Xi from the Dianping App Search Team for their valuable feedback on this paper. We also appreciate the strong support from the Meituan Research Collaboration Department.}}


\newpage

\bibliographystyle{ACM-Reference-Format}
\bibliography{sample-base}


\end{sloppypar}
\end{document}